\documentclass[aps,showpacs]{revtex4}
\usepackage{amsbsy}
\usepackage{graphicx,color}
\usepackage{amsfonts}
\usepackage{amsmath}
\usepackage{amssymb}

\newcommand{\ds}{ _{\downarrow}}
\newcommand{\us}{ _{\uparrow}}
\newcommand{\up}{\uparrow}
\newcommand{\down}{\downarrow}
\newcommand{\nn}{\nonumber}

\begin{document}
\draft
\title{The 3-body Coulomb problem}
\author{R. Combescot}
\address{Laboratoire de Physique Statistique, Ecole Normale Sup\'erieure, PSL Research University, 
UPMC Paris 06, Universit\'e Paris Diderot, CNRS, 24 rue Lhomond, 75005 Paris,
France.}
\date{Received \today}
\pacs{03.65.-w , 31.15.-p , 71.35.-y}

\begin{abstract}
We present a general approach for the solution of the three-body problem for a general interaction, and apply it to the case of the Coulomb interaction. 
This approach is exact, simple and fast. It makes use of integral equations derived from the consideration of the scattering properties of the system. 
In particular this makes full use of the solution of the two-body problem, the interaction appearing only through the corresponding known T-matrix.
In the case of the Coulomb potential we make use of a very convenient expression for the T-matrix obtained by Schwinger. 
As a check we apply this approach to the well-known problem of the
Helium atom ground state and obtain a perfect numerical agreement with the known result for the ground state energy. The wave function
is directly obtained from the corresponding solution. We expect our method to be in particular quite useful for the trion problem in semiconductors.
\end{abstract}
\maketitle

\section{INTRODUCTION}
Few body systems and problems \cite{blume,greene} are ubiquitous in almost all fields of physics, they arise for example in particle physics, 
nuclear physics, atomic physics, condensed matter physics, and so on. Since two-body problems are easily solved analytically or numerically,
the first level of non trivial problems arise with three-body problems. Among the first examples in quantum mechanics has been the Helium
atom, more specifically its ground state energy, where one deals with the Coulomb interaction. 
This has been addressed first by Hylleras \cite{hyll} by variational methods and pushed recently
to extraordinary precision \cite{jcp}. Another quite similar case is the H$^-$ ion \cite{frolov} which is remarkable for its very weakly bound
ground state and is of astrophysical interest \cite{chandra}. Yet another example is found in semiconductor physics, where the trion, i.e. a
bound state of an exciton and an electron (or a hole) \cite{lampert} is observed through its absorption or emission spectrum \cite{kheng,huard}.
This is again a case where the interaction is essentially the Coulomb interaction.
Three-body systems arise also because they may have their own intrinsic interest, 
such as the well known Efimov trimers \cite{greene,ferl}, with their remarkable
scaling properties, which have been the subject of much activity recently in nuclear physics and in cold atoms physics.

The recent surge of activity in ultracold gases \cite{pitstr,gps}, following the achievement of Bose-Einstein condensation in these systems,
has led to a renewed interest in few body physics \cite{blume,greene}. Indeed the situation is much simplified in these cases because these
systems are dilute and the relevant atomic energies are very low. As a result in most cases the interaction can be considered essentially as a contact interaction,
and in the scattering amplitude the contributions other than s-wave can be safely ignored. All the possible complexities of the interaction
potential disappear and the interaction is fully characterized by the scattering length. There is no dependence of the 
scattering amplitude on wavevectors, it depends only on energy. This makes the three-body Schr\"odinger equation much simpler
to solve since one has to deal with free atoms except for a boundary condition when two atoms are at the same position.
Similarly the scattering properties are much easier to find and for example the dimer-atom scattering length is obtained by solving
a one-dimensional integral equation, as initiated a long time ago by Skorniakov and Ter-Martirosian \cite{stm} for the neutron-deuteron problem.
These problems are very convenient to formulate in a diagrammatic formalism \cite{bevako} and to generalize to four-body problems
\cite{bkkcl}, leading again to fairly simple integral equations.

This diagrammatic approach has a further very attractive interest. Indeed it makes full use of the solution of the two-body problem.
Actually the interaction potential never appears explicitly in the equations, it comes in only through the two-body propagator
corresponding to the solution of the two-body problem. This looks a very reasonable and attractive approach to the solution
of the three-body problem: it makes much more sense to use the already known solution of the two-body problem rather than start
again from the beginning, as if the two-body problem had not been solved. This feature is so attractive that it is worthwhile
to explore if it can be extended with the same advantages to the case of a general interaction potential, getting rid of the
simplified contact interaction suited to cold gases. Actually this spirit is very close to another approach to the many-body problem,
the 'composite boson' formalism \cite{mon}, where the eigenstates of the two-body problem are taken as a new basis in which 
the whole many-body problem is rewritten. In this approach one makes again full use of the solution of the two-body problem.
In particular the trion problem has already been addressed within this approach \cite{sycmon}.

It is the purpose of the present paper to explore this generalization. We find that this extension can indeed been done with minimal
increase in complexity. As a result we find a new method to solve the three-body problem which is at the same time exact, simple and fast.
In practice, when we will come to explicit use, we will consider the specific case of the 3D Coulomb potential which is appropriate to
the case of the Helium ground state we will consider explicitly, and also to the case of the trion which we have mainly in mind.
This Coulomb potential case turns out to be particularly convenient since there is a simple analytic expression found by Schwinger \cite{schwing}
for the T-matrix, which sums up the solution of the two-body Coulomb problem. However there is no real problem to extend our
method to any interaction potential, and also to any dimension D. One has merely to obtain, analytically or numerically, the corresponding T-matrix
for the two-body problem.
An interpolation method can then be used for example to store the result for practical use in the numerical calculation.
Beyond providing an efficient way to solve any three-body problem, we hope that this approach can be extended
to the four-body problem along the same lines. But exploration of this path is naturally left for further work.
Such an extension would naturally be extremely useful for many problems, in particular to treat appropriately exciton-exciton
interaction in semiconductors which is of importance for the Bose-Einstein condensation of excitons \cite{ccd}.

The paper is organized as follows. In the next section \ref{short}, as an introduction, we will consider the case of cold gases,
where the interaction is short-ranged, and review the calculation
of the atom-dimer scattering length which contains the backbone of our procedure. Then in section \ref{gip} we generalize the approach
to a general interaction potential, leading to an integral equation for a 3-body scattering amplitude whose poles give the bound states
energies and the eigenfunctions of the 3-body problem. In the following section \ref{wf} it is shown explicitly how the wavefunction of
a bound state is obtained from the solution of the integral equation. We then specialize to the Coulomb potential and review in section
\ref{coulT} the derivation of the corresponding T-matrix by Schwinger. Finally we make use of our results in section \ref{Hegs}
to obtain the Helium atom ground state energy, which is found in perfect agreement with known results. We also give our results for
the ground state wavefunction. The last section is a summary and conclusion.

To summarize, the present paper is devoted to present our approach and to check it on a very well known case,
the Helium atom ground state. Its application to other interesting cases, in particular the energy of the trion, is left for further work.
For convenience and to be definite,  we prefer to adopt for our presentation
the semiconductor vocabulary specific to the case of the trion, since we have it in mind, rather than keep a general, vague and unspecific wording. 
Hence our three particles are one hole and two electrons, which have in
most of the paper opposite spins $\up$ and $\down$. The translation to any other physical situations of interest is obvious.

\section{Short-range interaction}\label{short}

Let us first recall what happens when we replace by a
short-range interaction the Coulomb interaction between the hole, with mass $m_h$, and the electrons,
with mass $m_e$, which we have mostly in mind. In addition we will omit in this section the interaction between the two electrons, 
and we will assume the simplest situation
where these electrons are identical (they have the same spin). This case is useful since this is the simplest one
in our class of problems.
This is basically the problem handled a long time ago by Skorniakov and Ter-Martirosian \cite{stm} to obtain the deuteron-neutron scattering length $a_3$.
This is also the situation found in cold gases. Here we treat it by making use of the diagrammatic method \cite{bevako}.

\begin{figure}
\centering
{\includegraphics[width=0.8\linewidth]{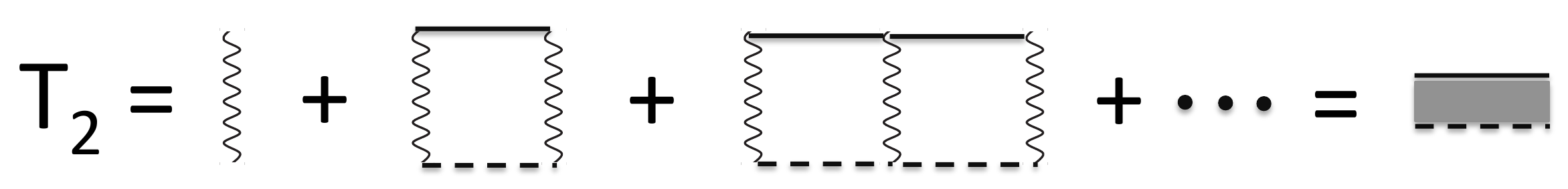}}
\caption{Exciton or dimer propagator. The full line is an electron propagator and the dashed line a hole propagator. The wavy line corresponds to
an interaction.}
\label{fig1T2}
\end{figure}

This case of the short-range interaction is quite simple because, at the low energy and wavevectors we are interested in, the dimer (or exciton) propagator does not 
depend on the entering or outgoing wavevectors but only on the total energy $\Omega $ and momentum ${\bf P}$ of the dimer. The situation is even simpler
since only the energy $\Omega_r $ of the relative motion is actually entering. It is given by $\Omega_r =\Omega - {\bf P}^2/2M$, where $M=m_e+m_h$
is the total mass of the particles making up the dimer (or exciton). Its explicit expression is, within a factor, the scattering amplitude for the relative motion and it
depends only on the scattering length $a$ which in this way is the only parameter necessary to characterize fully the interaction. Specifically it is given 
diagrammatically by Fig.\ref{fig1T2} and its explicit expression is:
\begin{eqnarray}\label{eqT2}
T_2(P) = \frac{2\pi}{\mu}\,\frac{1}{a^{-1} -\sqrt{2\mu (\mathbf{P}^2/2M-\Omega-i0_{+}})} 
\end{eqnarray}
where $P = \{\Omega,\textbf{P}\}$ is the momentum-energy four-vector, and $\mu =m_e m_h/(m_e+m_h)$ is the reduced mass. We have set $\hbar=1$
as we will do everywhere in the paper. 
The above expression has a single pole for $\Omega_r =-1/(2\mu a^2) \equiv -E_0$
corresponding to the single bound state of the two particles. This happens only when $a>0$, which we assume in the following, otherwise there is no
bound state and accordingly no dimer.

We want now to write an integral equation for the scattering amplitude of an electron on the exciton. Qualitatively this quantity is analogous to the $T_2$
considered just above, except that the hole is replaced by the exciton. In the case of short-range interaction the
situation with respect to the variables coming in the exciton-electron vertex $T_3$ is quite simple. Let us call $P$ the total momentum-energy of the
electron and the exciton, and $p$ the corresponding value for the entering electron. Hence the momentum-energy of the exciton is $P-p$, and we know from
Eq.(\ref{eqT2}) that this quantity is enough to fully characterize the entering exciton through its propagator. Similarly if the momentum-energy of the
outgoing electron is $\bar{p}$, the outgoing exciton has a momentum-energy $P-\bar{p}$. Accordingly $T_3$ depends only on $P$, $p$ and $\bar{p}$.
Note that, just as we have done above for $T_2(P)$ (see Fig.\ref{fig1T2}), we do not include in the expression for $T_3(p,\bar{p};P)$ the entering and
outgoing propagators for the electron and the exciton, since they would anyway be factored out in the equation we are looking for.

\begin{figure}
\centering
{\includegraphics[width=0.8\linewidth]{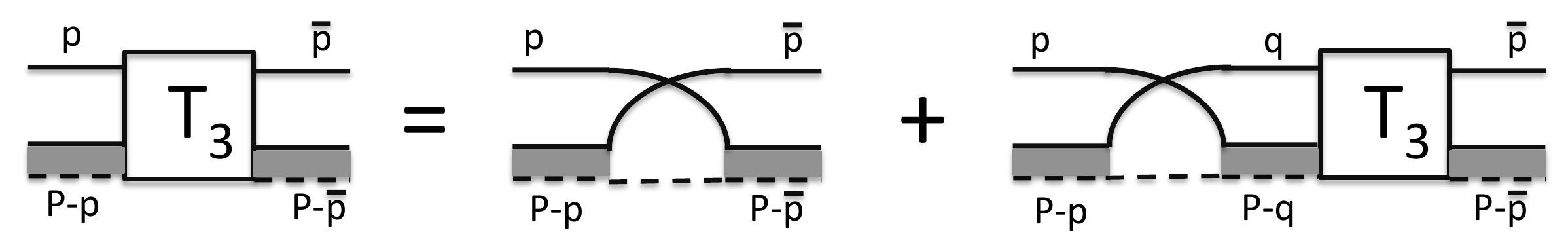}}
\caption{Diagrammatic representation of integral equation Eq.(\ref{eqintT3}) for the exciton-electron scattering vertex $T_3$. 
Full line: electron propagator. Dashed line: hole propagator. Shaded line: Exciton propagator}
\label{fig2T3}
\end{figure}

In order to obtain our equation we note that the simplest process arising in the electron-exciton scattering is merely the exchange of the incoming electron
with the one making up the exciton. More precisely, since all the interactions between the electron and the hole forming the entering exciton are already
taken into account in the entering exciton propagator, the only possible interaction which can arise in $T_3$ is between the hole of the exciton and the
incoming electron. However this interaction may be followed by another one between the same particles, and another one, and so on. Summing up
all these possible interactions gives an exciton propagator. The simplest case arises if this exciton propagator coincides with the outgoing exciton
propagator, as described by the first term in the right-hand side of Fig.~\ref{fig2T3}. We see that it corresponds indeed to an electron exchange.
However any other process may also occur with this exciton propagator and the electron before the final state. But the sum of all these processes
is precisely $T_3$ by definition. This is described by the second term in the right-hand side of Fig.~\ref{fig2T3}. This leads to the integral equation
represented diagrammatically in Fig.\ref{fig2T3}. It reads algebraically:
\begin{equation}\label{eqintT3}
T_3(p,\bar{p};P)=-g_h(P-p-\bar{p})- \sum_{q}g_h(P-p-q)g_e(q)\, T_2(P -q)\; T_3(q, \bar{p}; P)
\end{equation}
where, according to Feynmann diagrams rules, $\sum\limits_q\equiv i \int d{\bf q}\,d\omega_q/(2\pi)^4$. 

Here $g_h(p) \equiv g(\{\omega_p,{\bf p}\}) = 1/\left(\omega_p-\mathbf{p}^2/2m_h+i0_{+}\right)$ is the hole Green's function,
while $g_e(q) = 1/\left(\omega_q-\mathbf{q}^2/2m_e+i0_{+}\right)$ is the electron Green's function.
Finally the minus signs in the right-hand side of Eq.(\ref{eqintT3}) comes from the fact that we are exchanging the two electrons
and that this permutation of these two identical fermions implies a sign change.

We can integrate on the frequency $\omega_q$ in Eq.(\ref{eqintT3})
by closing the integration contour in the lower complex half-plane for the variable $\omega _q$.
Indeed from their definition $g_h(P-p-q)$ and $T_2(P -q)$ are analytical functions of $\omega_q$ in this domain.
Moreover, as given by Eq.(\ref{eqintT3}) itself $T_3(p, \bar{p}; P)$ is an analytical function of $\omega_p$ for
$\text{Im}\, \omega_p <0$, so $T_3(q, \bar{p}; P)$ is also analytical in the lower complex half-plane for the variable $\omega _q$.
Accordingly the only singularity in this region is the simple pole coming from $g_e(q)$.
Hence, by residue integration, only the on-the-shell value (\emph{i.e.} evaluated for $\omega _q={\bf q}^2/2m_e$) of the integrand comes in.
This leads us to consider the simpler problem of finding $T_3(\{{\bf p}^{2}/2m_e,{\bf p}\}, \bar{p}; P)$ by restricting $p$ to be also taken on-the-shell.

Moreover for the problems of physical interest, such as finding the scattering length or the ground state energy, we do not need to consider
general values for $\bar{p}$ and $P$. The scattering length corresponds to a situation where all the momenta in $T_3(p,\bar{p};P)$ go to zero
while the total energy is the exciton ground state energy, since the energy of the scattering electron goes to zero. Similarly we will find the
energy $-E$ (with $E>0$) of the bound states of the exciton-electron system, and in particular the ground state energy, by looking for resonances of the
exciton-electron scattering amplitude when all the momenta are zero. Hence we can restrict ourselves to the specific case $\bar{p}=0$
and ${\bf P}={\bf 0}$, so that $P=\{-E,{\bf 0}\}$. On the other hand we can not set from the start ${\bf p}={\bf 0}$ since in this case we could not write
an integral equation. We have to consider ${\bf p} \neq {\bf 0}$, and after having found the solution let possibly ${\bf p}$ go to zero.

As a result, setting $T_3(\{{\bf p}^{2}/2m_e,{\bf p}\}, 0; \{-E,{\bf 0}\})=t({\bf p})$ (we do not write explicitly the dependence on $E$) we obtain from Eq.(\ref{eqintT3})
the simpler integral equation
\begin{equation}\label{eqintT3fin}
t({\bf p})=\frac{2\mu}{2\mu E+{\bf p}^2}+\frac{2\mu }{(2\pi )^3}  \int \!d{\bf q} \;
\frac{T_2(\{-(E+{\bf q}^2/2m_e),{\bf q}\})}{2\mu E+{\bf p}^2+{\bf q}^2+2\mu\,{\bf p}.{\bf q}/m_h}\,t({\bf q})
\end{equation}
Actually $t({\bf p})$ depends only on the single variable $|{\bf p}|$, as it is obvious by rotational invariance, the angular integrations are easily performed
explicitly in the right-hand side, and we are left with a single variable integration. The integral equation is very easily solved numerically. For example for the scattering
length considered by Skorniakov and Ter-Martirosian \cite{stm}, all the fermion masses are equal $m_e=m_h=m$, and one has to set $E=E_0=1/ma^2$.
The scattering length $a_3$ is then related to the solution by $a_3=8\,t(0)/(3ma)$. One finds $a_3 = 1.18 \,a$.

\section{General interaction potential}\label{gip}

We will consider now a general interaction potential $V({\bf r})$, with ${\bf r}={\bf r}_e-{\bf r}_h$, 
and only later on specialize to the Coulomb interaction. Such a general case brings
immediately formal complications, which remain in the Coulomb case. They appear as soon as we consider the exciton propagator $T_2$.
Indeed the entering and outgoing wavevectors are relevant variables, whereas for the short-range interaction they are always small enough
compared to the cut-off wavevector to be taken equal to zero. This is clear when we notice that the two-body propagator is
directly related to the one-body propagator corresponding to the relative motion of the electron and the hole. Let us call ${\bf k}_e$ and $\omega_e$
the entering wavevector and energy of the electron, with similarly ${\bf k}_h$ and $\omega_h$ for the entering hole, together with
${\bf k}'_e$ and $\omega'_e$, and similarly ${\bf k}'_h$ and $\omega'_h$ for the outgoing electron and hole respectively. From momentum and energy
conservation we have ${\bf P} = {\bf k}_e + {\bf k}_h = {\bf k}'_e + {\bf k}'_h$, and $\Omega = \omega_e + \omega_h = \omega'_e + \omega'_h$.
For the relative motion the relevant energy is again $\Omega_r =\Omega - {\bf P}^2/2M$, while the entering momentum is 
${\bf k}=\left(m_h {\bf k}_e - m_e {\bf k}_h\right)/M={\bf k}_e-(m_e/M){\bf P}$ 
and the outgoing momentum is ${\bf k}'=\left(m_h {\bf k}'_e - m_e {\bf k}'_h\right)/M={\bf k}'_e-(m_e/M){\bf P}$.
For our purpose the relative motion problem is solved as soon as we have the corresponding Green's function $G(\omega,{\bf k},{\bf k}')$
defined by:
\begin{eqnarray}\label{}
G(\omega,{\bf k},{\bf k}')=\langle {\bf k}|\frac{1}{\omega - H}|{\bf k}'\rangle
\end{eqnarray}
where $H={\bf p}^2/2\mu + V({\bf r})$ is the Hamiltonian corresponding to the relative motion. This Green's function is basically obtained by
solving the Schr\"odinger equation for $H$.

Formally this Green's function satisfies the equation:
\begin{eqnarray}\label{eqG}
G=g+gVG
\end{eqnarray}
where $g(\omega,{\bf p}) = 1/\left(\omega-\mathbf{p}^2/2\mu +i0_{+}\right)$ is the free particle Green's function for the relative motion.
Since our $T_2$ is precisely the T-matrix for this relative motion, which satisfies:
\begin{eqnarray}\label{eqT2a}
T_2=V+VgT_2
\end{eqnarray}
it is easily obtained from $G$, since it is readily checked that, if $G$ satisfies Eq.(\ref{eqG}), then $T_2=g^{-1}Gg^{-1}-g^{-1}$ satisfies Eq.(\ref{eqT2a}).
This gives explicitly:
\begin{eqnarray}\label{eqT2G}
T_2(\omega,{\bf k},{\bf k}')=(\omega-\frac{\mathbf{k}^2}{2\mu})\Big(G(\omega,{\bf k},{\bf k}')-g(\omega,{\bf k})\delta_{{\bf k},{\bf k}'}\Big)
(\omega-\frac{\mathbf{k}'^2}{2\mu})
\end{eqnarray}
Hence we see that, while for short-range interaction, $T_2$ depends only on the four-vector $P = \{\Omega,\textbf{P}\}$, we have now to take also
into account the dependence on the incoming ${\bf k}$ and outgoing ${\bf k}'$ wavevectors for the relative motion. Accordingly we will denote it
$T_2(P;{\bf k},{\bf k}')$, the total energy $\Omega$ entering only, as above, through the combination $\Omega_r =\Omega - {\bf P}^2/2M$.

This complication implies a corresponding complication for $T_3$. Precedingly for short-range interaction we could specify only the total four-vector
of the exciton. Now we have also to specify the wavevectors of the electron and the hole making up the exciton after its break-up since they will enter
in the $T_2$ describing all the further processes. In practice it is not more complicated to start by specifying the four-vector corresponding to these two
particles. 

To be more specific we will now specialize to the case relevant for our problem of finding the ground state energy of our three-body system. Since,
for the actual cases we consider, it
will clearly be found for the two electrons having opposite spins, we will restrict ourselves to this case. Accordingly, in contrast with the preceding 
section where the two electrons were indistinguishable, they will now be distinguishable particles 
and in particular we will have no exchange processes.
This leads us to introduce two scattering vertices for the electron and the exciton, instead of a single one as in the preceding section. We will call $T_{3\up}$
the vertex corresponding to an electron $\up$ scattering on an exciton made of a hole and a $\down$ spin electron. Similarly we introduce $T_{3\down}$
corresponding to the situation where the electron spins are exchanged. We recall that, in $T_{3\up}$, since all the hole - $\down$ electron interactions 
have been taken into account in the incoming exciton propagator, the only possible interaction of the hole after the exciton break-up is with the $\up$ electron.
However there is another possibility, namely that the two electrons scatter. This possible process did not enter in the preceding section for cold gases
because, at very low energy, the scattering is dominantly s-wave which is forbidden for two identical fermions. In contrast we want naturally to take here into
account the Coulomb repulsion between electrons. 

Accordingly we will have to consider the T-matrix corresponding to the sum of the repeated scattering between the two electrons, which is analogous to
$T_2(P;{\bf k},{\bf k}')$ except that the interaction is now repulsive and there is naturally no bound state. We will denote this matrix by $T_2^e(P;{\bf k},{\bf k}')$.
Correspondingly we have to consider a $T_3$ vertex where the entering electrons have just interacted repeatedly so that they can not longer interact
and the first interaction to be considered is between the hole and one of these electrons. We denote this vertex $T_{3h}$. Ultimately the three vertices
$T_{3\up}$,$T_{3\down}$ and $T_{3h}$ are a way to describe all the possible scattering between the hole and the two electrons. On the other hand
we will have to consider for these three vertices only the case where the outgoing particles are the $\up$ electron and the exciton 
(just as the entering particles in $T_{3\up}$). As above we denote by $P$ the momentum-energy four-vector, and by ${\bar p}$ the momentum-energy of
the outgoing $\up$ electron. Actually, just as above, we will finally take ${\bar p}=0$ and $P=\{-E,{\bf 0}\}$. So for simplicity we will not indicate these
variables in the $T_3$ vertices. On the other hand as we have indicated above we have to indicate the momentum-energy of the three entering particles.
However since their sum is $P$, we need only to write it for two particles and we choose to write the variables for the two electrons. The first variable is for
the $\up$ electron and the second one for the $\down$ electron. Hence we have the three vertices $T_{3\up}(p,p')$,$T_{3\down}(p,p')$ and $T_{3h}(p,p')$.
In order to have notations similar to the ones used in section \ref{short} we include for example in the definition of $T_{3\up}(p,p')$ the free hole and
free down electron propagators corresponding to the broken exciton, and similarly for $T_{3\down}(p,p')$ and $T_{3h}(p,p')$.

\begin{figure}
\centering
{\includegraphics[width=0.8\columnwidth]{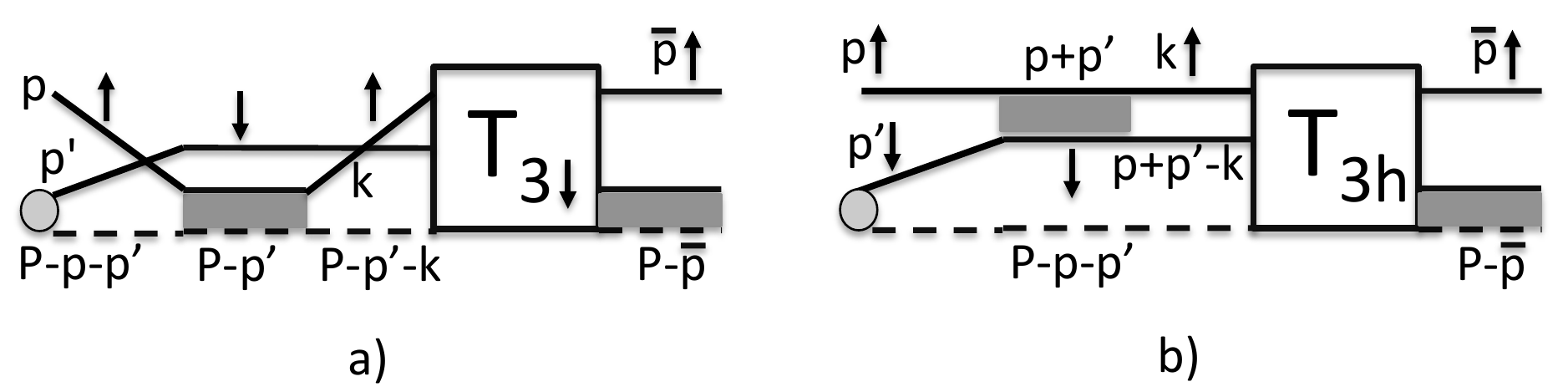}}
\caption{The two contributions to the integral equation Eq.(\ref{eqT3up}) for $T_{3\up}(p,p')$:
a) The $\up$ electron has its first interaction with the hole coming from the exciton.
b) The $\up$ electron has its first interaction with the $\down$ electron coming from the exciton.
Same notations as in Fig. \ref{fig2T3}. The shaded circle indicates that the hole and the $\down$ electron are coming
from the exciton, and that the first interaction in the diagram should not be between them.
$T_{3\down}$ and $T_{3h}$ are the corresponding vertices given respectively by Eq.(\ref{eqT3down}) and Eq.(\ref{eqT3hole}).
Note that there is no change of sign due to fermion lines crossings,
the one appearing on the figures are for readability and can be removed by deforming appropriately the propagator lines}
\label{fig3}
\end{figure}

Proceeding as for Eq.(\ref{eqintT3}) we can now write integral equations relating these vertices. Let us start with $T_{3\up}(p,p')$. In contrast to the
first term of Eq.(\ref{eqintT3}) there is no exchange term since the electrons have opposite spins. Hence, either the hole interacts with the $\up$ electron, 
which is described by $T_2$ and then any process may happen, which is described by $T_{3\down}$. 
This is shown diagrammatically in Fig. \ref{fig3} a). This is completely analogous to the second
term of Eq.(\ref{eqintT3}). However another possibility is that the two electrons interact, as described by $T_2^e$, 
followed by all the processes described by $T_{3h}$. This is represented diagrammatically in Fig. \ref{fig3} b). 
The corresponding equation for $T_{3\up}(p,p')$ gathering these two possible kinds of processes reads:
\begin{eqnarray}\label{eqT3up}
T_{3\up}(p,p')=g_e(p')g_h(P-p-p')\sum_{k}\Big[T_2(P -p';{\bf p}{-}r({\bf P}{-}{\bf p}')),{\bf k}{-}r({\bf P}{-}{\bf p}'))\; T_{3\down}(k,p') \\ \nn
+T_2^e(p+p';{\bf p}{-}\frac{{\bf p}{+}{\bf p}'}{2},{\bf k}{-}\frac{{\bf p}{+}{\bf p}'}{2})\; T_{3h}(k,p+p'-k)\Big]
\end{eqnarray}
where we have set $r=m_e/M$. For the case of electron-electron scattering this ratio becomes merely $1/2$.

\begin{figure}
\centering
{\includegraphics[width=0.8\columnwidth]{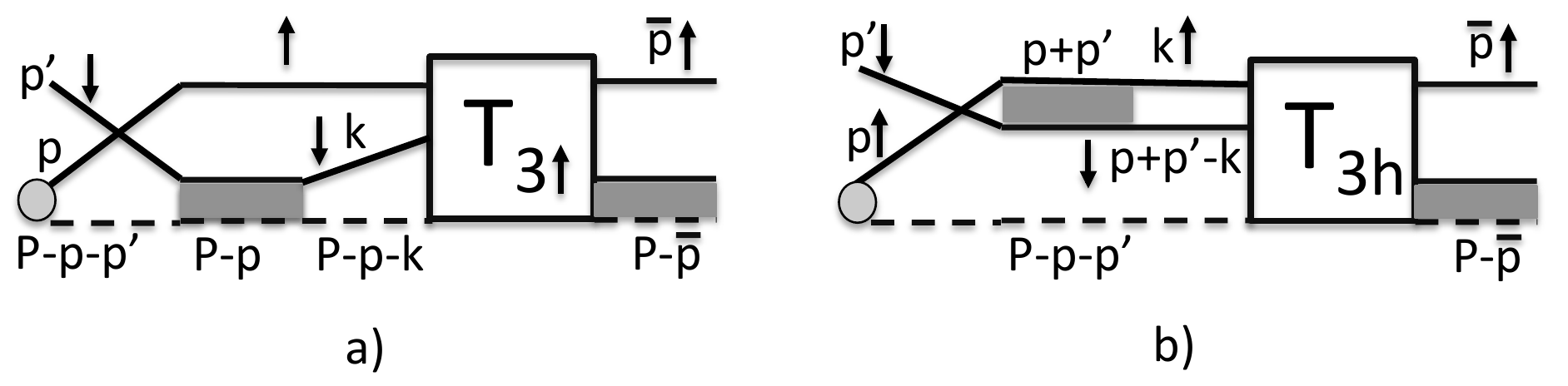}}
\caption{The two contributions to the integral equation Eq.(\ref{eqT3down}) for $T_{3\down}(p,p')$:
a) The $\down$ electron has its first interaction with the hole coming from the exciton.
b) The $\down$ electron has its first interaction with the $\up$ electron coming from the exciton.
Same notations as in Fig. \ref{fig2T3}. The shaded circle indicates that the hole and the $\up$ electron are coming
from the exciton, and that the first interaction in the diagram should not be between them.
$T_{3\up}$ and $T_{3h}$ are the corresponding vertices given respectively by Eq.(\ref{eqT3up}) and Eq.(\ref{eqT3hole}).}
\label{fig4}
\end{figure}

Similarly for $T_{3\down}(p,p')$ we have the possibility that the hole interacts with the $\down$ electron, described by $T_2$, followed by
all the processes corresponding to $T_{3\up}(p,p')$, as shown in Fig.\ref{fig4} a). 
There is also the possibility shown in Fig.\ref{fig4} b) of having the two electrons interacting as described by $T_2^e$,
followed by $T_{3h}$ processes. However there is finally a process, not possible for $T_{3\up}(p,p')$, which is merely that the incoming exciton breaks 
to produce the outgoing $\up$ electron while the incoming $\down$ electron forms with the hole the outgoing dimer. Since some involved propagators
are already factored out, this gives just an additional term $\delta_{p,0}\,g_h(P-p-p')$ if we take into account ${\bar p}=0$. This leads to the equation:
\begin{eqnarray}\label{eqT3down}
T_{3\down}(p,p')=\delta_{p,0}\,g_h(P-p-p')+g_e(p)g_h(P-p-p')\sum_{k}\Big[T_2(P -p;{\bf p'}{-}r({\bf P}{-}{\bf p}),{\bf k}{-}r({\bf P}{-}{\bf p}))\; T_{3\up}(p,k) \\ \nn
+T_2^e(p+p';{\bf p}{-}\frac{{\bf p}{+}{\bf p}'}{2},{\bf k}{-}\frac{{\bf p}{+}{\bf p}'}{2})\; T_{3h}(k,p+p'-k)\Big]
\end{eqnarray}

\begin{figure}
\centering
{\includegraphics[width=0.8\columnwidth]{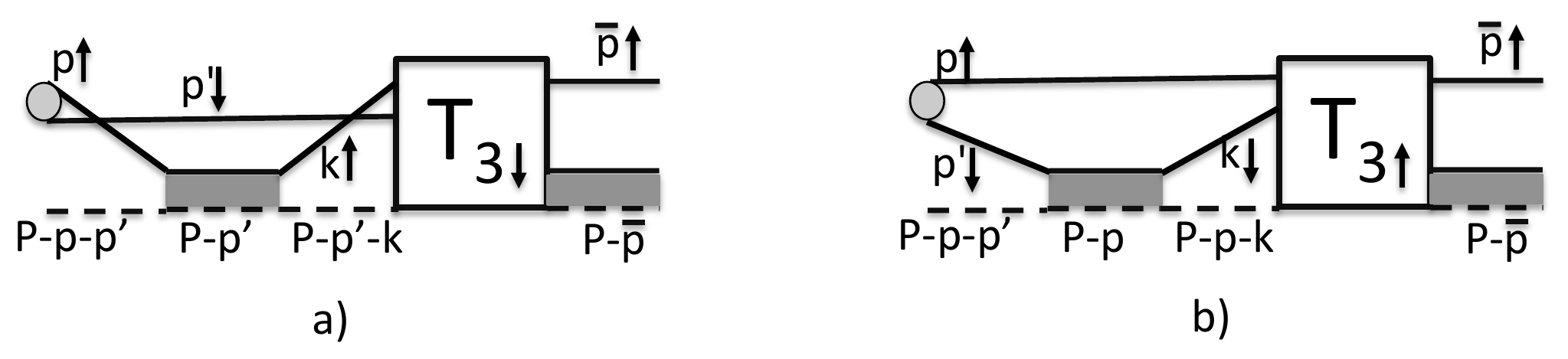}}
\caption{The two contributions to the integral equation Eq.(\ref{eqT3hole}) for $T_{3h}(p,p')$:
a) The hole has its first interaction with the $\up$ electron.
b) The hole has its first interaction with the $\down$ electron.
Same notations as in Fig. \ref{fig2T3}. The shaded circle indicates that the two electron had repeated interaction, 
and that the first interaction in the diagram should not be between them.
$T_{3\up}$ and $T_{3\down}$ are the corresponding vertices given respectively by Eq.(\ref{eqT3up}) and Eq.(\ref{eqT3down}).}
\label{fig5}
\end{figure}

Finally we have to write a similar equation for $T_{3h}(p,p')$. We have the possibilities that either one of the electron interacts with the hole,
which is described by $T_2$, followed by either $T_{3\up}(p,p')$ or $T_{3\down}(p,p')$, as shown on Fig.\ref{fig5} a) and b). 
But there is again the simple case where the $\down$
electron forms the outgoing exciton with the hole, the remaining $\up$ electron giving the outgoing electron. This leads as above to:
\begin{eqnarray}\label{eqT3hole}
T_{3h}(p,p')=\delta_{p,0}\,g_e(p')+g_e(p)g_e(p')\sum_{k}\Big[T_2(P -p;{\bf p'}{-}r({\bf P}{-}{\bf p})),{\bf k}{-}r({\bf P}{-}{\bf p}))\; T_{3\up}(p,k) \\ \nn
+T_2(P -p';{\bf p}{-}r({\bf P}{-}{\bf p}')),{\bf k}{-}r({\bf P}{-}{\bf p}'))\; T_{3\down}(k,p')\Big]
\end{eqnarray}

There is in our problem an additional symmetry which we have not used, namely the fact that the Hamiltonian is invariant with respect 
to the exchange of the two electrons. However it is appearant in our equations. We see that, if we write the equation for the combination
$T_{3+}(p,p') \equiv T_{3\up}(p,p')+T_{3\down}(p',p)$, it depends only on $T_{3+}$ in the right-hand side. Similarly the equation for
$T_{3-}(p,p') \equiv T_{3\up}(p,p')-T_{3\down}(p',p)$ depends only on $T_{3-}$. With respect to the $T_{3h}$ contribution, we can put it
under a more convenient form by making first, in Eq.(\ref{eqT3up}) and Eq.(\ref{eqT3down}), the change of variable $k \to k+(p+p')/2$,
and then, in the term coming from Eq.(\ref{eqT3down}), the change $k \to -k$. This leads to:
\begin{eqnarray}\label{eqT3pm}
T_{3\pm}(p,p')=\pm \delta_{p',0}\,g_h(P-p-p') + g_e(p')g_h(P-p-p')\sum_{k}\Big[\pm T_2(P -p';{\bf p}{+}r{\bf p}',{\bf k}{+}r{\bf p}')\; T_{3\pm}(p',k) \\ \nn
+T_2^e(p+p';\frac{{\bf p}{-}{\bf p}'}{2},{\bf k})\; T_{3h\pm}(\frac{p+p'}{2}{+}k,\frac{p+p'}{2}{-}k)\Big]
\end{eqnarray}
where we have set $T_{3h\pm}(p,p') \equiv T_{3h}(p,p') \pm T_{3h}(p',p)$. We have also used $T_2^e(p+p';-\frac{{\bf p}{-}{\bf p}'}{2},-{\bf k})=
T_2^e(p+p';\frac{{\bf p}{-}{\bf p}'}{2},{\bf k})$ valid for a potential satisfying $V(-{\bf r})=V({\bf r})$. Finally we have used the fact that we restrict
ourselves to the case ${\bf P}={\bf 0}$ to simplify the equation.

To close our set of equations we need an equation for $T_{3h\pm}(p,p')$ which is readily obtained from Eq.(\ref{eqT3hole}).
We find
\begin{eqnarray}\label{eqT3hpm}
T_{3h\pm}(p,p')=\delta_{p,0}\,g_e(p')+g_e(p)g_e(p')\sum_{k}T_2(P -p;{\bf p'}{+}r{\bf p},{\bf k}{+}r{\bf p})\; T_{3\pm}(p,k) \pm (p \leftrightarrow p')
\end{eqnarray}
Hence we have simplified our problem from three coupled equations for the three vertices $T_{3\up}$,$T_{3\down}$ and $T_{3h}$ to two coupled
equations for the vertices $T_{3\pm}$ and $T_{3h\pm}$. Naturally we can even introduce the expression for $T_{3h\pm}(p,p')$ into the equation
for $T_{3\pm}$ to obtain a single equation, but this is not particularly convenient. Clearly we will find the ground state in the symmetric subset
$T_{3+}$ and $T_{3h+}$, and accordingly we will restrict ourselves for simplicity to this case in the following, although the equations for the antisymmetric
subset can be similarly obtained with a few changes of sign.

A further simplification appears in the equations when we notice that, in the right-hand side of Eq.(\ref{eqT3pm}) and Eq.(\ref{eqT3hpm}),
only appear the vertices summed over the frequency component $\omega_k$ of the variable $k$, namely only $ \sum_{\omega_k}
T_{3+}(p',k)$ and $\sum_{\omega_k} T_{3h+}((p+p')/2{+}k,(p+p')/2{-}k)$. This could be expected since $T_2$ depends separately on the wavevectors
of the incoming and outgoing particles, but only on their total energy. Introducing the new functions $T_{3}(p,{\bf k}) \equiv 
\sum_{\omega_k} T_{3+}(p,k)$ and $S_{3}(p,{\bf k}) \equiv \sum_{\omega_k} T_{3h+}(p{+}k,p{-}k)=\sum_{\omega_k} T_{3h+}(p{-}k,p{+}k)$, 
we obtain the following equation for them by summing Eq.(\ref{eqT3pm}) over the frequency $\omega_{p'}$:
\begin{eqnarray}\label{eqT3red}
T_{3}(p,{\bf p}')=\delta_{{\bf p}',{\bf 0}}\,g_h(P-p) +  \sum_{\omega_{p'}} g_e(p')g_h(P-p-p')
\sum_{\bf k}\Big[T_2(P -p';{\bf p}{+}r{\bf p}',{\bf k}{+}r{\bf p}')\; T_{3}(p',{\bf k}) \\ \nn
+T_2^e(p+p';\frac{{\bf p}{-}{\bf p}'}{2},{\bf k})\; S_{3}(\frac{p+p'}{2},{\bf k})\Big]
\end{eqnarray}
Furthermore setting in Eq.(\ref{eqT3hpm}) $p=Q-q$ and $p'=Q+q$ and summing over the frequency $\omega_{q}$ we obtain for $S_3$
\begin{eqnarray}\label{eqT3hred}
S_{3}(\!Q,{\bf q}){=}\delta_{{\bf Q},{\bf q}}\,g_e(2Q){+}\!\sum_{\omega_q}g_e(Q{-}q)g_e(Q{+}q)\!\sum_{\bf k}T_2(P{-}Q{+}q;
{\bf Q}{+}{\bf q}{+}r({\bf Q}{-}{\bf q}),{\bf k}{+}r({\bf Q}{-}{\bf q}))T_{3}(\!Q{-}q,{\bf k}) \\ \nn
{+}({\bf q}\leftrightarrow\!-{\bf q})
\end{eqnarray}

We will now proceed to perform the frequencies integration. As in section \ref{short} this relies on the analytical properties of the various involved
quantities. To make the equations more transparent in this respect, let us rewrite for a moment Eq.(\ref{eqT3red}) and Eq.(\ref{eqT3hred}) by displaying only
the frequency variables. We have
\begin{eqnarray}\label{eqT3redfr}
T_{3}(\omega_p)=g_h(-E-\omega_p) +  \sum_{\omega_{p'}} g_e(\omega_{p'})g_h(-E-\omega_p-\omega_{p'})
\left[T_2(-E-\omega_{p'})\; T_{3}(\omega_{p'})+T_2^e(\omega_p+\omega_{p'})\; S_{3}(\frac{\omega_p+\omega_{p'}}{2})\right]
\end{eqnarray}
\begin{eqnarray}\label{eqT3hredfr}
S_{3}(\omega_Q)=g_e(2\omega_Q)+\sum_{\omega_q}g_e(\omega_Q-\omega_q)g_e(\omega_Q+\omega_q)
T_2(-E-\omega_Q+\omega_q)T_{3}(\omega_Q-\omega_q)+({\bf q}\leftrightarrow\!-{\bf q})
\end{eqnarray}
Making in the second term of the right-hand side of Eq.(\ref{eqT3redfr}) the change of variable $\omega_{p'}=\omega_x-\omega_p$,
and in Eq.(\ref{eqT3hredfr}) the change $\omega_q=\omega_Q-\omega_y$,
we find:
\begin{eqnarray}\label{eqT3redfrx}
T_{3}(\omega_p){=}g_h({-}E{-}\omega_p)+\sum_{\omega_{p'}} g_e(\omega_{p'})g_h({-}E{-}\omega_p{-}\omega_{p'})
T_2(-E{-}\omega_{p'})T_{3}(\omega_{p'}) \\ \nn
+\sum_{\omega_x}g_e(\omega_x{-}\omega_p)g_h({-}E{-}\omega_x)T_2^e(\omega_x)S_{3}(\frac{\omega_x}{2})
\end{eqnarray}
\begin{eqnarray}\label{eqT3hredfry}
S_{3}(\omega_Q)=g_e(2\omega_Q)+\sum_{\omega_y}g_e(\omega_y)g_e(2\omega_Q-\omega_y)
T_2(-E-\omega_y)T_{3}(\omega_y)+({\bf q}\leftrightarrow\!-{\bf q})
\end{eqnarray}
where naturally one should again understand $\sum_{\omega_x} \to i  \int d\omega_x/(2 \pi )$, and similarly for the frequency variables.

The second equation shows that $S_{3}(\omega_Q)$ is analytical in the upper $\omega_Q$ complex plane. On the other hand $T_{3}(\omega_p)$
is analytical in the lower $\omega_p$ complex plane, since the three terms in the right-hand side of Eq.(\ref{eqT3redfrx}) have this property. We can
now make use of these properties to perform the frequency integration. In Eq.(\ref{eqT3redfrx}), just as in Eq.(\ref{eqintT3}), we see that in the second
term in the right-hand side all the factors except $g_e(\omega_{p'})$ are analytical in the lower $\omega_{p'}$ complex plane. Closing the $\omega_{p'}$
integration contour by a semi-circle at infinity in this half-plane, the only contribution in a residue integration comes from the pole of $g_e(\omega_{p'})$
at $\omega_{p'}={\bf p}'^2/2m_e$. Hence we need in particular to evaluate $T_{3}(\omega_{p'})$ only  on-the-shell. Proceeding in the same way
in Eq.(\ref{eqT3hredfry}) we see once again that only the on-the-shell value of $T_{3}(y)$ appears. Finally, for the integration of the third term of
Eq.(\ref{eqT3redfrx}), we can close the contour in the upper-half $\omega_x$ complex plane where the only contribution comes from the pole of 
$g_h(-E{-}\omega_x)$
at $\omega_x=-E-{\bf x}^2/2m_h$, where ${\bf x}={\bf p}+{\bf p}'$, the other factors being analytical functions. So only the value of $S_{3}(\omega_x/2)$ for this specific frequency is required.

Coming back to the full equations Eq.(\ref{eqT3red}) and Eq.(\ref{eqT3hred}), we perform the same change of variables as described above
and perform the frequency integration in the way we have indicated. Since only on-the-shell quantities come in, this leads us to write the
equations for these quantities. We define $T({\bf p},{\bf p}')=T_{3}(\{{\bf p}^2/2m_e,{\bf p}\},{\bf p}')$ and 
$S({\bf Q},{\bf q})=S_{3}(\{-E-{\bf Q}^2/2m_h,{\bf Q}\}/2,{\bf q})$ and we find:
\begin{eqnarray}\label{eqT}
T({\bf p},{\bf p}')&=&-\frac{\delta_{{\bf p}',{\bf 0}}}{E+\frac{{\bf p}^2}{2\mu }}
-\frac{1}{E+\frac{{\bf p}^2+{\bf p}'^2}{2m_e}+\frac{({\bf p}+{\bf p}')^2}{2m_h}} \\  \nn
&\times &\!\!\sum_{\bf k}\Bigg[T_2\!\left(\!\{-E{-}\frac{{\bf p}'^2}{2m_e},{-}{\bf p}'\};{\bf p}{+}r{\bf p}',{\bf k}{+}r{\bf p}'\right)\;T({\bf p}',{\bf k})
+T_2^e\!\left(\!\{-E{-}\frac{({\bf p}{+}{\bf p}')^2}{2m_h},{\bf p}{+}{\bf p}'\};\frac{{\bf p}{-}{\bf p}'}{2},{\bf k}\right)\;S({\bf p}+{\bf p}',{\bf k})\Bigg]
\end{eqnarray}
\begin{eqnarray}\label{eqS}
S({\bf Q},{\bf q})&=&-\frac{\delta_{{\bf Q}/2,{\bf q}}}{E+\frac{{\bf Q}^2}{2\mu }}
-\frac{1}{E+\frac{{\bf Q}^2}{2m_h}+\frac{{\bf Q}^2+4{\bf q}^2}{4m_e}}  \\  \nn
&\times &\sum_{\bf k}\;T_2\left(\{-E{-}\frac{(\frac{{\bf Q}}{2}+{\bf q})^2}{2m_e},{-}(\frac{{\bf Q}}{2}+{\bf q})\};
(\frac{{\bf Q}}{2}{-}{\bf q}){+}r(\frac{{\bf Q}}{2}{+}{\bf q}),{\bf k}{+}r(\frac{{\bf Q}}{2}{+}{\bf q})\right)\;T(\frac{{\bf Q}}{2}{+}{\bf q},{\bf k})
+({\bf q}\leftrightarrow\!-{\bf q})
\end{eqnarray}
where we have used the explicit form of the propagators $g_e$ and $g_h$.

Although it does not look so simple, this set of equations is clearly the best we could hope for this problem. We have two vertices $T$ and $S$
instead of one because we take into account not only electron-hole interaction but also electron-electron interaction. Moreover these quantities
depend only on three variables, the modulus of each vector and the angle between them, which is expected since $T_2$ and $T_2^e$ depend
on the entering and outgoing wavevectors in the general case we are dealing with.

It is interesting to see how these equations simplify to something similar to what we had in section \ref{short} when $T_2$ depends only on the
total momentum-energy and its dependence 
on wavevectors can be neglected, as it is the case for the short-range interaction considered in the above section. Indeed in this case we see
that, in the right-hand sides of these equations, the summation over ${\bf k}$ introduces merely $ \sum_{\bf k}T({\bf p}',{\bf k})$,
$\sum_{\bf k}S({\bf p}+{\bf p}',{\bf k})$ and $\sum_{\bf k}T({\bf Q}/2{+}{\bf q},{\bf k})$. Introducing $t({\bf p})=\sum_{\bf k}T({\bf p},{\bf k})$ and
$s({\bf Q})=\sum_{\bf k}S({\bf Q},{\bf k})$, and summing Eq.(\ref{eqT}) and Eq.(\ref{eqS}) over ${\bf p}'$ and ${\bf q}$ respectively, we see
that we obtain a set of integral equations for $t({\bf p})$ and $s({\bf Q})$ which is quite simple, since $t({\bf p})$ and $s({\bf Q})$ depend actually
only on the single variables $|{\bf p}|$ and $|{\bf Q}|$ respectively, and the angular integrations can be performed easily. We do not write them
explicitly in the general case since we will not make use of them. 

However we will pursue this investigation in a quite particular case because it offers a simple and interesting check of our method.
First we restrict ourselves to the case where there is no electron-electron interaction, which means we take $T_2^e=0$.  This makes 
$s({\bf Q})$ irrelevant, and we have only to consider $t({\bf p})$. 
We obtain easily for $t({\bf p})$ an equation which is essentially identical to Eq.(\ref{eqintT3fin}), except for the signs because the electrons have now
opposite spins. We write it explicitly only in the additional particular case where the hole mass is infinite $m_h \to \infty$, so that $\mu =m_e$ and $r=0$.
This case is particularly simple because, in Eq.(\ref{eqT}), the dependence on the angle between ${\bf p}$ and ${\bf p}'$ disappears 
which makes the summation over ${\bf p}'$ easier.
We find, with now the simpler notation $|{\bf p}| \equiv p$:
\begin{eqnarray}\label{eqmingf}
t(p)=-\frac{1}{E+\frac{p^2}{2m_e }}-\frac{1}{2\pi ^2}  \int_{0}^{\infty} dp' \,p'^2\,\frac{t_2(-E{-}\frac{p'^2}{2m_e})}{E+\frac{p^2+p'^2}{2m_e}}\;t(p')
\end{eqnarray}
Here we have made explicit the fact that $T_2(\{\Omega,{\bf P}\})$ depends on the total momentum-energy $\Omega $ and ${\bf P}$ 
only through the relative motion energy $\Omega_r =\Omega - {\bf P}^2/2M$ by setting $T_2(\{\Omega,{\bf P}\}) \equiv t_2(\Omega_r)$. 
In our case $M=\infty$, which leads to $T_2(\{-E{-}p'^2/2m_e,{-}{\bf p}'\})= t_2(-E{-}p'^2/2m_e)$.
If we further specialize to the short-range situation we have from Eq.(\ref{eqT2}) $t_2(\Omega)=(2\pi a/m_e)[a^{-1}-\sqrt{-2m_e \Omega }]^{-1}$.

Let us now focus on the specific problem of this paper, namely finding the ground state energy of the three-body problem. This is obtained in
the general case by making use of the fact that our three-body vertices diverge when the energy is equal to a bound state energy, just in the
same way as $T_2$ has poles when the energy is equal to a bound state energy. This implies that, when $E$ is equal to the ground state
energy, the homogeneous parts of Eq.(\ref{eqT}) and Eq.(\ref{eqS}) have a solution. On the other hand, in the very particular case considered
just above, the ground state energy is obvious. Indeed, since the hole mass is infinite it can be considered as a fixed impurity, and the electrons 
just feel the attractive potential of this impurity. Moreover since they do not interact, we have just two independent one-body problem, one for
each electron. Hence the ground state energy is merely the sum of the ground state energy of each electron. In particular for the short-range
interaction, the ground state energy is twice the energy $E_0=1/2m_ea^2$ of the bound state. So the homogeneous part of Eq.(\ref{eqmingf})
should have a solution for $E=2E_0=1/m_ea^2$. We can make in this equation changes of function and variables appropriate to get rid of
$m_e$ and $a$. But this is equivalent to take $m_e$ and $a$ as units of mass and length. This leads us to conclude that the homogeneous
integral equation:
\begin{eqnarray}\label{eqintreduite}
t(p)=\frac{2}{\pi}  \int_{0}^{\infty} dp' \frac{p'^2}{p^2+p'^2+2}\,\frac{\sqrt{p'^2+2}+1}{p'^2+1}\;t(p')
\end{eqnarray}
should have a solution. Although the physical problem and the corresponding ground state energy are trivial, this is not the case for the
corresponding integral equation Eq.(\ref{eqintreduite}). Nevertheless it is easily checked that this equation has the solution
$t(p)=1/(\sqrt{p^2+2}+1)$, which we will derive more systematically below. Hence we have checked that our method gives the correct
ground state energy for this very particular case. Let us just mention that this check can be extended to the case where the two electrons
have different masses, where a similar but somewhat more involved solution can be found.

It is interesting to generalize the above check to the case of a general interaction. We consider again the case where the hole mass is
infinite and the electrons do not interact $T_2^e=0$, so the ground state energy is again twice the ground state of an electron in the presence of
the hole interaction potential $E=2E_0$. We start from our general equations Eq.(\ref{eqT}) and Eq.(\ref{eqS}), but again $S$ is irrelevant
and we are only left with Eq.(\ref{eqT}). As above the relative motion energy entering $T_2$ is $-(E+p'^2/2m_e)=-(2E_0+p'^2/2m_e)$, and
we have to make $\mu =m_e$, $m_h=\infty$ and $r=0$ in Eq.(\ref{eqT}). This leads us for the homogeneous integral equation to:
\begin{eqnarray}\label{eqTminf}
T({\bf p},{\bf p}')=
-\frac{1}{E+\frac{{\bf p}^2+{\bf p}'^2}{2m_e}}
\sum_{\bf k}T_2(-E{-}\frac{{\bf p}'^2}{2m_e},{\bf p},{\bf k})\;T({\bf p}',{\bf k})
\end{eqnarray}
with the notation $T_2(\{\Omega,{\bf P}\};{\bf p},{\bf k}) \equiv T_2(\Omega_r,{\bf p},{\bf k})$. The general expression of $T_2$ is obtained
from Eq.(\ref{eqT2G}) with $\omega =-(E+{\bf p}'^2/2m_e)$:
\begin{eqnarray}\label{eqT2Gbis}
T_2(\omega,{\bf k},{\bf k}')=(\omega-\frac{\mathbf{k}^2}{2m_e}) \sum_{n}\frac{\varphi_n ({\bf k})\,\varphi_n ({\bf k}')}{\omega -{\mathcal E}_n}
(\omega-\frac{\mathbf{k}'^2}{2m_e}) -(\omega-\frac{\mathbf{k}^2}{2m_e})\delta_{{\bf k},{\bf k}'}
\end{eqnarray}
where we have expressed the Green's function in terms of the eigenfunctions and eigenenergies of the relative motion hamiltonian
$H\varphi_n({\bf k})={\mathcal E}_n\varphi_n({\bf k})$. By time reversal invariance we can take the eigenfunctions as real.
Since in our specific case $E=2E_0$ only the ground state is involved, we may suspect
that only the ground state wavefunction $\varphi_0({\bf k})$ appears in $T({\bf p},{\bf p}')$. Indeed, making use of the orthonormality
of the eigenfunctions $ \sum_{\bf k}\varphi_m ({\bf k})\,\varphi_n ({\bf k})=\delta_{m,n}$, we find that:
\begin{eqnarray}\label{eqsolT}
T({\bf p},{\bf p}')=\frac{E_0+\frac{{\bf p}^2}{2m_e}}{2E_0+\frac{{\bf p}^2+{\bf p}'^2}{2m_e}} \varphi_0 ({\bf p})\,\varphi_0 ({\bf p}')
\end{eqnarray}
is solution as it is easily checked by carrying this expression into Eq.(\ref{eqTminf}) with $E=2E_0$.

Actually, for this case of a general interaction, we can extend the above argument to the case where the electrons are respectively
in excited states $\varphi_1 ({\bf p})$ and $\varphi_2 ({\bf p})$, with energy $-E_1$ and $-E_2$. Since they do not interact we should
find an excited bound state of our three-body system with energy $-(E_1+E_2)$. This should also give rise to a divergence of
our three-body vertex, so we should have a corresponding solution for Eq.(\ref{eqTminf}) for $E=E_1+E_2$. Indeed we have
found that this equation has the solution:
\begin{eqnarray}\label{eqsolT1}
T({\bf p},{\bf p}')=\frac{\left(E_1+\frac{{\bf p}^2}{2m_e}\right)\varphi_1 ({\bf p})\,\varphi_2 ({\bf p}')+
\left(E_2+\frac{{\bf p}^2}{2m_e}\right)\varphi_1 ({\bf p}')\,\varphi_2 ({\bf p})}{E_1+E_2+\frac{{\bf p}^2+{\bf p}'^2}{2m_e}}
\end{eqnarray}
which generalizes Eq.(\ref{eqsolT}).

Finally coming back to the short-range interaction case, we can immediately have the expression for the eigenfunction of the single bound state
by comparing the general expression Eq.(\ref{eqT2G}) of $T_2$ in the vicinity of the pole $\omega =-1/(2m_e a^2)=-E_0$ with its specific
expression given below Eq.(\ref{eqmingf}). This gives $\varphi_0 ({\bf p})=(8\pi /a)^{1/2}[p^2+a^{-2}]^{-1}$. We can then make use of the
general expression for the solution Eq.(\ref{eqsolT}) and of the definition $t({\bf p})=\sum_{\bf k}T({\bf p},{\bf k})$ to find the solution of
Eq.(\ref{eqintreduite}). In this way one recovers (for $a=1$, $m_e=1$) the solution $t({\bf p})=1/(1+\sqrt{{\bf p}^2+2})$ already given below Eq.(\ref{eqintreduite}).

\section{The wave function}\label{wf}

As we have explained above, the ground state energy of our 3-body problem (or actually any eigenenergy) will be obtained in the general case 
by requiring that the homogeneous part of Eq.(\ref{eqT}) and Eq.(\ref{eqS}) have a solution. Let us rewrite these homogeneous equations for clarity
\begin{eqnarray}\label{eqThom}
T({\bf p},{\bf p}')&=&
-\frac{1}{E+\frac{{\bf p}^2+{\bf p}'^2}{2m_e}+\frac{({\bf p}+{\bf p}')^2}{2m_h}} \\  \nn
&\times &\!\!\sum_{\bf k}\Bigg[T_2\!\left(\!\{-E{-}\frac{{\bf p}'^2}{2m_e},{-}{\bf p}'\};{\bf p}{+}r{\bf p}',{\bf k}{+}r{\bf p}'\right)\;T({\bf p}',{\bf k})
+T_2^e\!\left(\!\{-E{-}\frac{({\bf p}{+}{\bf p}')^2}{2m_h},{\bf p}{+}{\bf p}'\};\frac{{\bf p}{-}{\bf p}'}{2},{\bf k}\right)\;S({\bf p}+{\bf p}',{\bf k})\Bigg]
\end{eqnarray}
\begin{eqnarray}\label{eqShom}
S({\bf Q},{\bf q})&=&
-\frac{1}{E+\frac{{\bf Q}^2}{2m_h}+\frac{{\bf Q}^2+4{\bf q}^2}{4m_e}}  \\  \nn
&\times &\sum_{\bf k}\;T_2\left(\{-E{-}\frac{(\frac{{\bf Q}}{2}+{\bf q})^2}{2m_e},{-}(\frac{{\bf Q}}{2}+{\bf q})\};
(\frac{{\bf Q}}{2}{-}{\bf q}){+}r(\frac{{\bf Q}}{2}{+}{\bf q}),{\bf k}{+}r(\frac{{\bf Q}}{2}{+}{\bf q})\right)\;T(\frac{{\bf Q}}{2}{+}{\bf q},{\bf k})
+({\bf q}\leftrightarrow\!-{\bf q})
\end{eqnarray}

Once we have found the ground state energy of our 3-body problem, we may suspect that the corresponding wavefunction is related to the corresponding
residue of our $T_3$ matrix. Actually we have to handle carefully the frequency variables in order to find the proper relation between the wavefunction and the
residue. Moreover we have to take into account properly the fact that the matrices we have used above do not correspond precisely to the full $T_3$ matrix.

Let us define a Green's function ${\bar G}_3$ for the propagation of our three-body system. Precisely we set:
\begin{eqnarray}\label{}
{\bar G}_3(t,{\bf k}\us,{\bf k}\ds,{\bf k}_h,{\bf k'}\us,{\bf k'}\ds,{\bf k'}_h)=-i \langle 0 | c_{{\bf k}\us}(t)c_{{\bf k}\ds}(t)c_{{\bf k}_h}(t) c^{\dag}_{{\bf k'}_h}(0)c^{\dag}_{{\bf k'}\ds}(0)c^{\dag}_{{\bf k'}\us}(0) | 0 \rangle
\end{eqnarray}
where the operators $c_{{\bf k}\us,\ds,_h}(t)$ annihilate at time $t>0$ the $\up,\down$ electrons and the $h$ hole created at time $t=0$ by the operators
$c^{\dag}_{{\bf k}\us,\ds,_h}(0)$ acting on vacuum $| 0 \rangle$. If we introduce the eigenstates $| n\rangle$ and eigenenergies ${\mathcal E}_n$ of the
three-body Hamiltonian $H$, related by $H|n\rangle={\mathcal E}_n|n\rangle$, we have for the Fourier transform $G_3(\omega)$ of ${\bar G}_3(t)$:
\begin{eqnarray}\label{}
G_3(\omega,{\bf k}\us,{\bf k}\ds,{\bf k}_h,{\bf k'}\us,{\bf k'}\ds,{\bf k'}_h)= \sum_{n} \frac{\Phi_n({\bf k}\us,{\bf k}\ds)\Phi_n^*({\bf k'}\us,{\bf k'}\ds)}{\omega - {\mathcal E}_n + i0_{+} }
\end{eqnarray}
where in the wavefunction $\Phi_n({\bf k}\us,{\bf k}\ds)=\langle 0 | c_{{\bf k}\us}c_{{\bf k}\ds}c_{{\bf k}_h}|n\rangle$ we have taken into account that
the total momentum of our 3-body system is zero, which implies ${\bf k}_h=-({\bf k}\us+{\bf k}\ds)$. We see indeed that the residue corresponding to the
pole ${\mathcal E}_n$ is the product of the wavefunctions $\Phi_n({\bf k}\us,{\bf k}\ds)\Phi_n^*({\bf k'}\us,{\bf k'}\ds)$.

On the other hand we define the $T_3$ matrix with three different times, instead of a single one, by setting:
\begin{eqnarray}\label{}
{\bar {\mathcal T}}_3(t_1,t_2,t_3,{\bf k}\us,{\bf k}\ds,{\bf k}_h,{\bf k'}\us,{\bf k'}\ds,{\bf k'}_h)=(-i)^3 \langle 0 | c_{{\bf k}\us}(t_1)c_{{\bf k}\ds}(t_2)c_{{\bf k}_h}(t_3) c^{\dag}_{{\bf k'}_h}(0)c^{\dag}_{{\bf k'}\ds}(0)c^{\dag}_{{\bf k'}\us}(0) | 0 \rangle
\end{eqnarray}
Its Fourier transform ${\mathcal T}_3(\omega _1,\omega _2,\omega _3,{\bf k}\us,{\bf k}\ds,{\bf k}_h,{\bf k'}\us,{\bf k'}\ds,{\bf k'}_h)$ with respect to $t_1,t_2,t_3$
is basically the quantity we have dealt with from the beginning of the paper. Note however that, for simplicity, we do not try to describe properly the final state of our
3-body problem since when we consider the pole contribution at ${\mathcal E}_n$, the entering and the outgoing variables decouple completely. The variables we
consider correspond actually to the entering variables of the scattering problem.

We recover ${\bar G}_3(t)$ by taking $t_1=t_2=t_3=t$ in ${\mathcal T}_3(t_1,t_2,t_3)$. In Fourier transform this implies the relation:
\begin{eqnarray}\label{}
G_3(\omega)=(\frac{i}{2\pi })^2  \int (\prod_{i} d\omega_i )\delta(\omega - \sum_{i} \omega_i)
{\mathcal T}_3(\omega _1,\omega _2,\omega _3)=
(\frac{i}{2\pi })^2  \int d\omega_1 d\omega_2 {\mathcal T}_3(\omega _1,\omega _2,\omega -\omega _1-\omega _2)
\end{eqnarray}
where, for clarity, we have not written the momentum variables. This last relation is easily checked in the particular case where there is no interaction
between the particles in our 3-body system. One finds as expected that the 3-body wavefunction is merely the product of the wavefunctions of each particle.

Switching back to our original variables, and taking again into account that the total momentum is zero, this means that we will obtain the 3-body wavefunction
corresponding to the energy ${\mathcal E}_n$ as the residue at $\Omega = {\mathcal E}_n$ of:
\begin{eqnarray}\label{int3}
(\frac{i}{2\pi })^2  \int d\omega_p d\omega_{p'} {\mathcal T}_3(p,p',P-p-p')
\end{eqnarray}
with $P = \{\Omega,\textbf{P}\}$ and ${\bf P}={\bf 0}$.

\begin{figure}
\centering
{\includegraphics[width=0.8\linewidth]{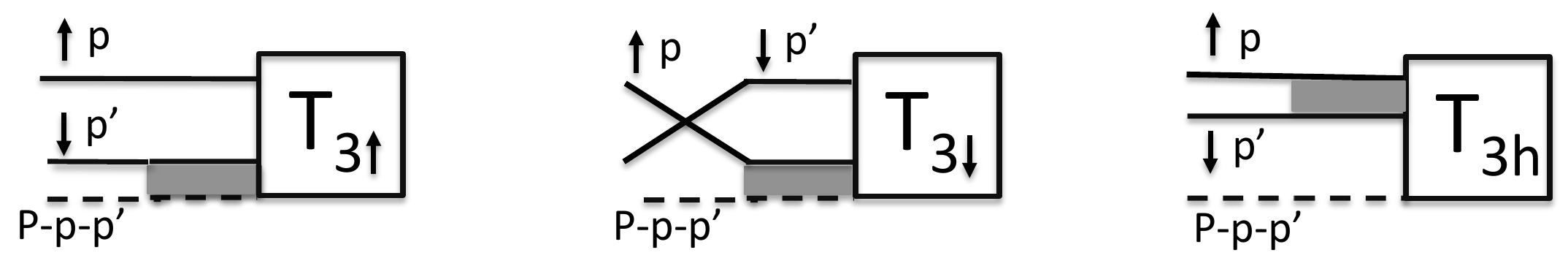}}
\caption{The three contributions to the expression Eq.(\ref{eqT3gen}) for the connected part of the general vertex ${\mathcal T}_3$.}
\label{figT3}
\end{figure}

We have now to write ${\mathcal T}_3$ in terms of the vertices $T_{3\up}$,$T_{3\down}$ and $T_{3h}$ which we have introduced earlier in section \ref{gip}.
Naturally in doing this we disregard all the disconnected diagrams which do not contribute to the ground state we are looking for. In the definition of $T_{3\up}$,$T_{3\down}$ and $T_{3h}$, we had not included the free propagators of the incoming particles as well as the $T_2$ matrix corresponding to the first two interacting particles. We have now to write them explicitly to obtain ${\mathcal T}_3$, as it is shown in Fig.\ref{figT3}.

This leads to
\begin{align}\label{eqT3gen}
{\mathcal T}_3(p,p',P-p-p')&=g_e(p)g_e(p')g_h(P-p-p')\Big[\sum_{p'_1}T_2(P -p;{\bf p}'{+}r{\bf p},{\bf p}_1'{+}r{\bf p})\; T_{3\up}(p,p_1') \\ \nn
+\sum_{p_1}T_2(P -p';\,&{\bf p}{+}r{\bf p}',{\bf p}_1{+}r{\bf p}')\; T_{3\down}(p_1,p')
+\sum_{k}T_2^e(p+p';{\bf p}{-}\frac{1}{2}({\bf p}+{\bf p}'),{\bf k}{-}\frac{1}{2}({\bf p}+{\bf p}'))\; T_{3h}(k,p+p'-k)\Big]
\end{align}
where we have taken ${\bf P}={\bf 0}$ into account. We have then introduced $T_{3\pm}(p,p')= T_{3\up}(p,p')\pm T_{3\down}(p',p)$
and we are looking for a pole where $T_{3+}(p,p')$ diverges while $T_{3-}(p,p')$ does not. This implies $2 T_{3\up}(p,p')= 2 T_{3\down}(p',p)=T_{3+}(p,p')$
in the vicinity of the pole. Similarly $T_{3h\pm}(p,p') = T_{3h}(p,p') \pm T_{3h}(p',p)$ where $T_{3h+}(p,p')$ diverges while  $T_{3h-}(p,p')$ does not, which leads to $2 T_{3h}(p,p')= 2 T_{3h}(p',p)=T_{3h+}(p,p')$. This leads to
\begin{align}\label{eqt3}
2{\mathcal T}_3(p,p',P-p-p')&=g_e(p)g_e(p')g_h(P-p-p')\sum_{k}\Big[T_2(P -p;{\bf p}'{+}r{\bf p},{\bf k}{+}r{\bf p})\; T_{3+}(p,k) \\ \nn
+T_2(P -p';\,&{\bf p}{+}r{\bf p}',{\bf k}{+}r{\bf p}')\; T_{3+}(p',k)
+T_2^e(p+p';\frac{1}{2}({\bf p}-{\bf p}'),{\bf k})\; T_{3h}(\frac{1}{2}(p+p')+k,\frac{1}{2}(p+p')-k)\Big]
\end{align}
where we have made the change $k \to k+(p+p')/2$ in the last term. We now make use of Eq.(\ref{eqT3pm}) and Eq.(\ref{eqT3hpm}), where the first term in the right-hand side is omitted since it is not divergent. When we calculate from
these equations the combination $g_e(p)T_{3+}(p,p')+g_e(p')T_{3+}(p',p)+g_h(P-p-p')T_{3h+}(p,p')$, making appropriately the change of variable $k \to -k$ and using of $T_{3h+}(p,p')=T_{3h+}(p',p)$, we obtain twice the right-hand side of Eq.(\ref{eqt3}). This leads to
\begin{eqnarray}\label{}
4{\mathcal T}_3(p,p',P-p-p')=g_e(p)T_{3+}(p,p')+g_e(p')T_{3+}(p',p)+g_h(P-p-p')T_{3h+}(p,p')
\end{eqnarray}
To obtain the wavefunction we still have from (\ref{int3}) to sum this quantity over $\omega_p$ and $\omega_p'$. However from the definitions
$T_{3}(p,{\bf k})= \sum_{\omega_k} T_{3+}(p,k)$ and $S_{3}(p,{\bf k}) = \sum_{\omega_k} T_{3h+}(p{+}k,p{-}k)=\sum_{\omega_k} T_{3h+}(p{-}k,p{+}k)$, given above Eq.(\ref{eqT3red}), we have:
\begin{eqnarray}\label{}
\sum_{\omega_p,\omega_p'} g_e(p)T_{3+}(p,p')=\sum_{\omega_p}g_e(p)T_{3}(p,{\bf p}')\hspace{20mm} \sum_{\omega_p,\omega_p'} g_e(p')T_{3+}(p',p)=\sum_{\omega_p'}g_e(p')T_{3}(p',{\bf p})
\end{eqnarray}
and
\begin{eqnarray}\label{}
\sum_{\omega_p,\omega_p'} g_h(P-p-p')T_{3h+}(p,p')=\sum_{\omega_q}g_h(P-q)S_{3}(\frac{q}{2},{\bf k})
\end{eqnarray}
where this last relation has been obtained by going from the variables $p$ and $p'$ to the variables $q=p+p'$ and
$k=(p-p')/2$, and using the definition of $S_{3}(q/2,{\bf k})$.

Hence we are left with the calculation of
\begin{eqnarray}\label{evT3fin}
4\sum_{\omega_p,\omega_p'}{\mathcal T}_3(p,p',P-p-p')=\sum_{\omega_p}g_e(p)T_{3}(p,{\bf p}')+\sum_{\omega_p'}g_e(p')T_{3}(p',{\bf p})
+\sum_{\omega_q}g_h(P-q)S_{3}(\frac{q}{2},\frac{{\bf p}-{\bf p}'}{2})
\end{eqnarray}
We have seen (see Eq.(\ref{eqT3redfrx})) that $T_{3}(p,{\bf p}')$ is analytical for Im$\,\omega_p < 0$. Going from summation to integration by
$\sum_{\omega_p} \to i  \int d\omega_p/(2 \pi )$, and closing the integration contour on $\omega_p$ in the lower complex plane, only the pole
from $g_e(p)$ contributes, and the result is just the on-the-shell value of $T_{3}(p,{\bf p}')$, namely $T_{3}(\{{\bf p}^2/2m_e,{\bf p}\},{\bf p}')=T({\bf p},{\bf p}')$.
Similarly the second term in Eq.(\ref{evT3fin}) gives $T({\bf p}',{\bf p})$. Finally (see Eq.(\ref{eqT3hredfry})) $S_{3}(q/2,({\bf p}-{\bf p}')/2)$ is analytical for 
Im$\,\omega_q > 0$. Closing the integration contour in the $\omega_q$ upper complex plane, we obtain from the definition $S({\bf Q},{\bf q})=S_{3}(\{-E-{\bf Q}^2/2m_h,{\bf Q}\}/2,{\bf q})$ that the third term in Eq.(\ref{evT3fin}) is $S({\bf p}+{\bf p}',\frac{{\bf p}-{\bf p}'}{2})$.

Finally we end up with the conclusion that the 3-body bound state wavefunction is given, from the solution $T({\bf p},{\bf p}')$ and $S({\bf Q},{\bf q})$ 
of Eq.(\ref{eqThom}) and Eq.(\ref{eqShom}), by:
\begin{eqnarray}\label{eqwav}
T({\bf p},{\bf p}')+T({\bf p}',{\bf p})+S({\bf p}+{\bf p}',\frac{{\bf p}-{\bf p}'}{2})
\end{eqnarray}
within a multiplicative constant, since the solutions $T$ and $S$ of the homogeneous equations are not normalized. This expression is, as
expected, invariant under the exchange of ${\bf p}$ and ${\bf p}'$.

Note that, when we make the substitution ${\bf Q}={\bf p}+{\bf p}'$ and ${\bf q}=({\bf p}-{\bf p}')/2$ in Eq.(\ref{eqShom}) for $S({\bf Q},{\bf q})$,
we obtain an expression which is identical to the $T_2$ term (that is the first term in the right-hand side) in Eq.(\ref{eqThom}) for 
$T({\bf p},{\bf p}')+T({\bf p}',{\bf p})$. Hence there is no need, in calculating the wavefunction Eq.(\ref{eqwav}), to evaluate Eq.(\ref{eqShom})
since this is already done when Eq.(\ref{eqThom}) is evaluated. In particular, if we come back to the case where the two electrons are non interacting,
in which case $T_2^e=0$, the wavefunction from Eq.(\ref{eqwav}) is $2[T({\bf p},{\bf p}')+T({\bf p}',{\bf p})]$. We obtain the same result (without the
irrelevant factor of 2) if we argue that in this case $S$ is irrelevant from Eq.(\ref{eqThom}) and that only the $T$ terms should contribute in Eq.(\ref{eqwav}).
From the explicit solution Eq.(\ref{eqsolT}) which we have found in this case, we see that the ground state wavefunction is just the product 
$\varphi_0 ({\bf p})\,\varphi_0 ({\bf p}')$ of the two single electron ground state wavefunctions, as expected.

\section{The Coulomb T-matrix}\label{coulT}

Let us now come to our specific problem of handling the Coulomb potential. Since the $T_2$ is directly linked to the Green's function, itself obtained
by solving the Schr\"odinger equation, one expects to be able to write the Coulomb $T_2$ in terms of solutions of the Coulomb Schr\"odinger equation,
namely hypergeometric functions. This can indeed be done, but this leads to expressions which are not so easy to handle numerically.
On the other hand it is obviously quite important to have a convenient expression for this $T_2$ in order to obtain a numerically efficient
solution for our problem, which is one of our basic purpose. Fortunately such an expression has been obtained by Schwinger \cite{schwing}.
Since it is not so well known, let us review briefly its derivation for completeness. This will also allows us to set our notations.

The simplicity of the result is linked to the hidden symmetry of the Coulomb potential, which gives rise in classical mechanics to the existence
of a special conserved quantity, the Lenz vector, and which has been used by Pauli in its operatorial solution of the Coulomb Schr\"odinger
equation.

The hamiltonian is:
\begin{eqnarray}\label{eqHCoul}
H=-\frac{1}{2\mu }\Delta_{\bf r}-\epsilon_s \frac{Ze^2}{4\pi \epsilon r}
\end{eqnarray}
where $e$ is the electronic charge, $\epsilon$ is the permittivity of the medium, be it vacuum or semiconductor. In case of the inter-electronic 
repulsion we have $\epsilon_s=-1$ and $Z=1$, while for electron-hole attraction $\epsilon_s=1$ and $Z$ depends on the "hole" charge, 
since we want also to consider the case of He where we will have $Z=2$ for the nucleus.

It is convenient to take half the Bohr radius $a_0=4\pi \epsilon/(2\mu Z e^2)$ as unit of length. In the same way we take $1/(2\mu a_0^2)$ as energy unit
and we set $\omega =-\kappa^2/(2\mu a_0^2)$, where $\kappa$ will be real since we will actually have to consider only negative values for $\omega$.
Similarly we express the wavevectors in terms of the unit $1/a_0$, which leads to introduce reduced wavevectors by ${\bf k}={\bf q}/a_0$,
and so on for other wavevectors.
The equation:
\begin{eqnarray}\label{}
 \sum_{\bf k''}\langle {\bf k}|(\omega - H) |{\bf k}''\rangle G(\omega,{\bf k}'',{\bf k}')=\delta_{{\bf k},{\bf k}'}
\end{eqnarray}
for the Green's function becomes:
\begin{eqnarray}\label{eqGfCoulomb}
-(\kappa^2+{\bf q}^2)g(\kappa,{\bf q},{\bf q}')+\epsilon_s \frac{1}{2\pi ^2} \int \!d{\bf q}'' \,\frac{g(\kappa,{\bf q''},{\bf q}')}{({\bf q}-{\bf q}'')^2}
=(2\pi )^3 \delta({\bf q}-{\bf q}')
\end{eqnarray}
where we have introduced a reduced Green's function $g$ by $G(\omega,{\bf q}/a_0,{\bf q}'/a_0)=(2\mu a^5_0 )g(\kappa,{\bf q},{\bf q}')$,
and we have gone from discrete to continuous variables by $\sum\limits_q\to \int d{\bf q}/(2\pi)^3$ and $\delta_{{\bf q},{\bf q}'} \to 
(2\pi )^3 \delta({\bf q}-{\bf q}')$.

\begin{figure}
\centering
{\includegraphics[width=0.5\linewidth]{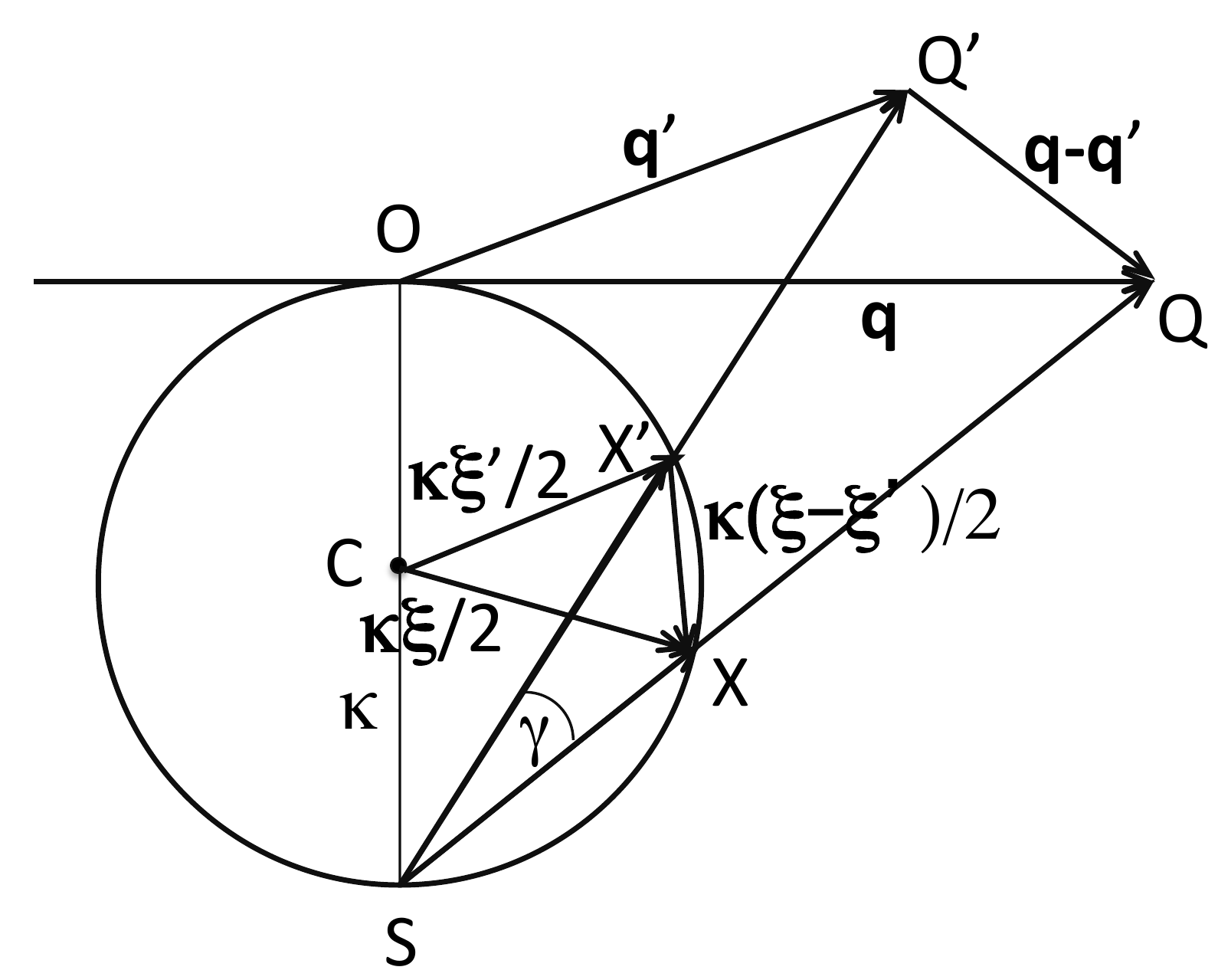}}
\caption{Schematic view of the change of variables from ${\bf q}$ to ${\boldsymbol \xi}$.}
\label{figstereo}
\end{figure}

We proceed now to a change of variables. Let O be the origin of the three-dimensional ${\bf q}$ space, and Q the point with ${\bf OQ}={\bf q}$.
We consider Q as the stereographic projection of a point X which is on a 4D-sphere with diameter $\kappa $, the 3D ${\bf q}$ space being tangent
to this sphere at the origin O, and the pole S of the stereographic projection being such that OS is a diameter of the sphere, see Fig.~\ref{figstereo}.
From elementary geometry we have $SX.SQ=\kappa ^2$. If C is the center of the sphere, we set ${\bf CX}=\frac{\kappa }{2}\,{\boldsymbol \xi}$, where
 ${\boldsymbol \xi}$ runs on a 4D sphere with unit radius ${\boldsymbol \xi}^2=1$ and therefore carries the information on the orientation of ${\bf CX}$. Our change of variables
 is from ${\bf q}$ to ${\boldsymbol \xi}$.
 
The components of the 4D vector ${\boldsymbol \xi}$ respectively parallel and perpendicular to the ${\bf q}$ plane are given by:
\begin{eqnarray}\label{}
{\boldsymbol \xi}_{\parallel}=\frac{2\kappa}{\kappa ^2+{\bf q}^2}\; {\bf q}\hspace{10mm}\xi_{\perp}=\frac{\kappa ^2-{\bf q}^2}{\kappa ^2+{\bf q}^2}
\end{eqnarray}  
If Q and Q' are two points in the ${\bf q}$ plane, with corresponding points X and X' on the sphere, we have in the triangle with sides SXQ and SX'Q',
and common angle $\gamma$ between these two sides, $QQ'^2=SQ^2+SQ'^2-2SQ.SQ' \cos \gamma$ and $XX'^2=SX^2+SX'^2-2SX.SX' \cos \gamma$.
Together with $SX.SQ=SX'.SQ'=\kappa ^2$, this leads to $XX'^2=\kappa^4 QQ'^2/(SQ^2.SQ'^2)$, that is:
\begin{eqnarray}\label{eqxiq}
({\boldsymbol \xi}-{\boldsymbol \xi}')^2=\frac{4\kappa ^2}{(\kappa ^2+{\bf q}^2)(\kappa ^2+{\bf q}'^2)}\;({\bf q}-{\bf q}')^2
\end{eqnarray}
Letting ${\bf q}' \to {\bf q}$ and ${\boldsymbol \xi}' \to {\boldsymbol \xi}$, this implies:
\begin{eqnarray}\label{}
d\xi=\frac{2\kappa }{\kappa ^2+{\bf q}^2}\,dq
\end{eqnarray}
for the corresponding infinitesimal lengths $dq$ and $d\xi$. Taking three such corresponding infinitesimal orthogonal variations for ${\bf q}$ and ${\boldsymbol \xi}$, this leads to the relation between elementary volumes:
\begin{eqnarray}\label{}
d{\boldsymbol \xi} = \left(\frac{2\kappa }{\kappa ^2+{\bf q}^2}\right)^3\,d{\bf q}
\end{eqnarray}
where $d{\boldsymbol \xi}$ is the elementary solid angle on the 4D unit sphere. This relation implies that the Jacobian of our transformation is 
$[2\kappa/(\kappa ^2+{\bf q}^2)]^3$. This has to be used in making the change of variables in the right-hand side of Eq.(\ref{eqGfCoulomb}):
\begin{eqnarray}\label{eqdeltaxiq}
\delta({\bf q}-{\bf q}')=\left(\frac{2\kappa }{\kappa ^2+{\bf q}^2}\right)^3\delta({\boldsymbol \xi}-{\boldsymbol \xi}')
\end{eqnarray}
Finally we make in Eq.(\ref{eqGfCoulomb}) the change:
\begin{eqnarray}\label{eqchgmtG}
g(\kappa,{\bf q},{\bf q}')=-\frac{(2\pi )^3(2\kappa)^3}{(\kappa ^2+{\bf q}^2)^2(\kappa ^2+{\bf q}'^2)^2}\;\Gamma(\kappa ,{\boldsymbol \xi},{\boldsymbol \xi}')
\end{eqnarray}
which leads to:
\begin{eqnarray}\label{eqGr4D}
\Gamma(\kappa ,{\boldsymbol \xi},{\boldsymbol \xi}')-\frac{\epsilon_s}{4\pi ^2\kappa} \int d{\boldsymbol \xi}''
\;\frac{\Gamma(\kappa ,{\boldsymbol \xi}'',{\boldsymbol \xi}')}
{({\boldsymbol \xi}-{\boldsymbol \xi}'')^2} = \delta({\boldsymbol \xi}-{\boldsymbol \xi}')
\end{eqnarray}

The integral is just a convolution product, and this equation is simply solved by
an expansion in spherical harmonics. These are the 4D generalization \cite{harmpolD} $Y_{n,j}({\boldsymbol \xi})$ of the standard 3D
spherical harmonics $Y_{lm}$. Actually since all the quantities in Eq.(\ref{eqGr4D}) depend only on the angle between the two involved directions,
an expansion in the corresponding \cite{harmpolD} 4D Legendre polynomials $P_n({\boldsymbol \xi}.{\boldsymbol \xi'})$ is enough.
They are related to the spherical harmonics by:
\begin{eqnarray}\label{}
P_n({\boldsymbol \xi}.{\boldsymbol \xi'})=\frac{S_{D-1}}{N(D,n)} \sum_{j=1}^{N(D,n)} Y_{n,j}({\boldsymbol \xi})Y^{*}_{n,j}({\boldsymbol \xi'})
\end{eqnarray}
Here $S_{D-1}$ is the surface of the unit sphere in dimension D ($S_2=4\pi $ and $S_3=2\pi ^2$), $j$ is collectively for the azimuthal numbers
necessary for the enumeration of the spherical harmonics with a given $n$, and $N(D,n)=[(2n+D-2)/n] (_{n-1}^{n+D-3})$ is the degeneracy of 
the $n$ level, that is the number of different spherical harmonics with a given $n$. For example $N(3,n)=2n+1$ and $N(4,n)=(n+1)^2$. The spherical harmonics
are orthonormal:
\begin{eqnarray}\label{}
 \int d{\boldsymbol \xi}\; Y^{*}_{n,j}({\boldsymbol \xi})Y_{n',j'}({\boldsymbol \xi})= \delta_{nn'} \delta_{jj'}
\end{eqnarray}
which implies:
\begin{eqnarray}\label{eqorthP}
 \int d{\boldsymbol \xi''}\; P_{n}({\boldsymbol \xi}.{\boldsymbol \xi}'')P_{n'}({\boldsymbol \xi}''.{\boldsymbol \xi}')=
 \frac{S_{D-1}}{N(D,n)} P_{n}({\boldsymbol \xi}.{\boldsymbol \xi}') \delta_{nn'}
\end{eqnarray}
Moreover they satisfy the closure relation:
\begin{eqnarray}\label{eqclos}
 \sum_{n,j} Y_{n,j}({\boldsymbol \xi})Y^{*}_{n,j}({\boldsymbol \xi'})=\sum_{n} \frac{N(D,n)}{S_{D-1}} P_{n}({\boldsymbol \xi}.{\boldsymbol \xi'})
 =\delta({\boldsymbol \xi}-{\boldsymbol \xi}')
\end{eqnarray}
Finally the Legendre polynomials are linked to their generating function by:
\begin{eqnarray}\label{eqfuncgen}
\frac{1}{(1-2rt+r^2)^{D/2-1}}= \sum_{n=0}^{\infty}\frac{D-2}{2n+D-2}N(D,n)\,r^n P_n(t)
\end{eqnarray}
which implies \cite{harmpolD}:
\begin{eqnarray}\label{}
\frac{1-r^2}{(1-2rt+r^2)^{D/2}}= \sum_{n=0}^{\infty}N(D,n)\,r^n P_n(t)
\end{eqnarray}
obtained by multiplying Eq.(\ref{eqfuncgen}) by $r^{D/2-1}$ and taking the derivative with respect to $r$.

Taking $D=4$ and $r=1$ in Eq.(\ref{eqfuncgen}), together with $({\boldsymbol \xi}-{\boldsymbol \xi}')^2=2(1-{\boldsymbol \xi}.{\boldsymbol \xi'})$, we have:
\begin{eqnarray}\label{eqpotCoul}
\frac{1}{({\boldsymbol \xi}-{\boldsymbol \xi}')^2}=\sum_{n=0}^{\infty}(n+1)\,P_n({\boldsymbol \xi}.{\boldsymbol \xi'})
\end{eqnarray}
Inserting a Legendre polynomial expansion of $\Gamma(\kappa ,{\boldsymbol \xi},{\boldsymbol \xi}')$ in Eq.(\ref{eqGr4D}):
\begin{eqnarray}\label{eqexpGam}
\Gamma(\kappa ,{\boldsymbol \xi},{\boldsymbol \xi}')= \sum_{n,j}
\Gamma_n(\kappa)Y_{n,j}({\boldsymbol \xi})Y^{*}_{n,j}({\boldsymbol \xi'})=
\sum_{n}\frac{N(4,n)}{S_{3}}
\Gamma_n(\kappa) P_n({\boldsymbol \xi}.{\boldsymbol \xi'})
\end{eqnarray}
together with Eq.(\ref{eqpotCoul}), Eq.(\ref{eqorthP}) and Eq.(\ref{eqclos}), we obtain the solution of Eq.(\ref{eqGr4D}) as:
\begin{eqnarray}\label{}
\Gamma_n(\kappa)=\frac{1}{1-\frac{\epsilon_s}{2\kappa (n+1)}}
\end{eqnarray}
In particular, in the attractive case $\epsilon_s=1$, the poles of $\Gamma_n(\kappa)$ are found for $2\kappa =1/(n+1)$ with $n=0,1,\cdots$
which give the expected energies $\omega =-\kappa^2/(2\mu a_0^2)=-\mu (Ze^2)^2/[2(n+1)^2(4\pi \epsilon )^2]$ for the bound states, 
with the expected degeneracy $N(4,n)=(n+1)^2$.

It is possible to sum up the series Eq.(\ref{eqexpGam}) for $\Gamma(\kappa ,{\boldsymbol \xi},{\boldsymbol \xi}')$ leading to a very nice closed
form. We write:
\begin{eqnarray}\label{eqgamdec}
\Gamma_n(\kappa)=\frac{1}{1-\frac{\epsilon_s}{2\kappa(n+1)}}=1+\frac{\epsilon_s}{2\kappa(n+1)}
+\frac{1}{(2\kappa )^2(n+1)}\frac{1}{n+1-\frac{\epsilon_s}{2\kappa}}
\end{eqnarray}
and make use in the last term of the integral representation:
\begin{eqnarray}\label{}
\frac{1}{n+1-\frac{\epsilon_s}{2\kappa}}= \int_{0}^{1}du\,u^{-\frac{\epsilon_s}{2\kappa}}u^n
\end{eqnarray}
Moreover, writing $1-2{\boldsymbol \xi}.{\boldsymbol \xi'} u+u^2=(1-u)^2+u({\boldsymbol \xi}-{\boldsymbol \xi}')^2$, we use Eq.(\ref{eqfuncgen}) to sum up
the corresponding series:
\begin{eqnarray}\label{}
 \sum_{n=0}^{\infty}\frac{u^n}{n+1}N(4,n)\,P_n({\boldsymbol \xi}.{\boldsymbol \xi'})=\frac{1}{(1-u)^2+u({\boldsymbol \xi}-{\boldsymbol \xi}')^2}
\end{eqnarray}
The first two terms in Eq.(\ref{eqgamdec}) are summed through Eq.(\ref{eqclos}) and Eq.(\ref{eqpotCoul}). This leads us finally to:
\begin{eqnarray}\label{eqfinGamma}
\Gamma(\kappa ,{\boldsymbol \xi},{\boldsymbol \xi}')=\delta({\boldsymbol \xi}-{\boldsymbol \xi}')+
\frac{\epsilon_s}{4\pi ^2 \kappa }\frac{1}{({\boldsymbol \xi}-{\boldsymbol \xi}')^2}+
\frac{1}{8\pi ^2 \kappa ^2}\int_{0}^{1}du\,\frac{u^{-\frac{\epsilon_s}{2\kappa}}}{(1-u)^2+u({\boldsymbol \xi}-{\boldsymbol \xi}')^2}
\end{eqnarray}
As pointed out by Schwinger, it is possible to change the integration contour to obtain, for any value of the reduced energy $\kappa $,
a well defined expression. However in deforming the contour, care must be taken of the possible contributions of the poles coming from the denominator.
But we will not need to perform such a transformation.

Let us now come to the T-matrix itself. It is convenient to introduce also reduced units in Eq.(\ref{eqT2G}). Introducing, just as for the Green's function,
a reduced T-matrix by $T_2(\omega,{\bf q}/a_0,{\bf q}'/a_0)=(a_0/2\mu)t_2(\kappa,{\bf q},{\bf q}')$, Eq.(\ref{eqT2G}) becomes:
\begin{eqnarray}\label{eqt2}
t_2(\kappa,{\bf q},{\bf q}')=(\kappa ^2+{\bf q}^2)(\kappa ^2+{\bf q}'^2)\left[g(\kappa,{\bf q},{\bf q}')-g_0(\kappa,{\bf q},{\bf q}')\right]
\end{eqnarray}
where $g_0(\kappa,{\bf q},{\bf q}')=-(2\pi )^3 \delta({\bf q}-{\bf q}')/(\kappa^2+{\bf q}^2)$. On the other hand we rewrite Eq.(\ref{eqfinGamma}) in terms
of the variable ${\bf q}$ by making use of Eq.(\ref{eqxiq}), Eq.(\ref{eqdeltaxiq}) and Eq.(\ref{eqchgmtG}). As could be expected, the 
$\delta({\boldsymbol \xi}-{\boldsymbol \xi}')$ term in Eq.(\ref{eqfinGamma}) cancels exactly with the $g_0$ term in Eq.(\ref{eqt2}), and we get finally:
\begin{eqnarray}\label{eqt2Coul}
t_2(\kappa,{\bf q},{\bf q}')=-\frac{4\pi \epsilon _s}{({\bf q}-{\bf q}')^2}-\frac{2\pi }{\kappa}\frac{1}{({\bf q}-{\bf q}')^2}\,
I(\kappa ,z)
\end{eqnarray}
where we have set:
\begin{eqnarray}\label{}
I(\kappa ,z)=\int_{0}^{1}du\,\frac{u^{-\frac{\epsilon_s}{2\kappa}}}{u+z(1-u)^2}\hspace{10mm}z=\frac{(\kappa ^2+{\bf q}^2)(\kappa ^2+{\bf q}'^2)}{4\kappa ^2 ({\bf q}-{\bf q}')^2}
\end{eqnarray}
The first term in Eq.(\ref{eqt2Coul}) is merely the well-known Born approximation for $t_2$, the only term to survive in the limit $\kappa \to \infty$.
Integrating by parts in $I(\kappa ,z)$, Eq.(\ref{eqt2Coul}) can also be conveniently rewritten as
\begin{eqnarray}\label{eqt2Coulbis}
t_2(\kappa,{\bf q},{\bf q}')=-\frac{4\pi \epsilon _s z}{({\bf q}-{\bf q}')^2}
\int_{0}^{1}du\,\frac{u^{-\frac{\epsilon_s}{2\kappa}}(1-u^2)}{[u+z(1-u)^2]^2}
\end{eqnarray}

In our domain of interest we have $z \ge 1/4$, the minimum being reached when ${\bf q}$ and ${\bf q}'$ are antiparallel with $|{\bf q}|.|{\bf q}|'=\kappa ^2$.
This implies that the poles $u_1$ and $u_2$ in the integrand of $I(\kappa ,z)$ are always complex conjugate. They are on the unit circle and they go from
$u_1=u_2=-1$ for $z=1/4$ to $u_1=u_2=1$ for $z \to \infty$. Hence they do not make any problem in the numerical evaluation of $I(\kappa ,z)$. It is
possible to express in general $I(\kappa ,z)$ in terms of hypergeometric functions, but this is not particularly useful for its numerical evaluation.
A particular case is for $z=1/4$ where one finds $I(\kappa ,1/4)=-2-(\epsilon _s/\kappa)[\psi(1/2-\epsilon _s/4\kappa)-\psi(-\epsilon _s/4\kappa)]$,
with $\psi(x)$ being the digamma function. 

In the case $\epsilon_s=1$ the discrete spectrum for $1/2\kappa =n$ gives rise to poles in $t_2$, which appear from 
$I(\kappa ,z)$ through the divergent behaviour of the integrand 
for $u \to 0$. In particular, for $\kappa  \to 1/2$, one finds easily $I(\kappa ,z) \simeq 1/[(2\kappa -1)z]$, leading to $t_2(\kappa,{\bf q},{\bf q}') \simeq
-4\pi /[(2\kappa -1)({\bf q}^2+1/4)({\bf q}'^2+1/4)]$. This has to be compared to the expression of the Green's function, which in reduced units, gives
$g(\kappa,{\bf q},{\bf q}') \simeq t_2(\kappa,{\bf q},{\bf q}')/[(\kappa ^2+{\bf q}^2)(\kappa ^2+{\bf q}'^2)] \simeq 2\pi /[(-\kappa^2+1/4)
({\bf q}^2+1/4)^2({\bf q}'^2+1/4)^2]$. This agrees with the general expression $\varphi_0 ({\bf k})\,\varphi_0 ({\bf k}')/(\omega -{\mathcal E}_0)$
of the Green's function in the vicinity of this pole, which reads in reduced units $\Phi_0 ({\bf q})\,\Phi_0 ({\bf q}')/(-\kappa ^2 +1/4)$,
and with the expression of the normalized ground state wavefunction $\Phi_0 ({\bf q})=(2\pi)^{1/2}/({\bf q}^2+1/4)^2$.

Let us finally note that it is possible to check analytically, with this expression of the ground state wavefunction and the expression of $t_2$ given
by Eq.(\ref{eqt2Coul}), that Eq.(\ref{eqsolT}) is indeed solution of Eq.(\ref{eqTminf}), although the corresponding calculation is not that simple.

\section{The case of the Helium ground state}\label{Hegs}

Let us now come to the explicit treatment of the Helium atom ground state (with naturally only Coulomb interaction between particles retained in the Hamiltonian).
Our basic purpose is to see how our method works in practice numerically and in particular to check numerically that it is exact.
However since the nucleus mass is very large compared to the electronic mass, we will simplify a bit our practical task by taking it infinite, i.e. $m_h=\infty$.
Since we only want to display an effective application of our method, this is an unimportant simplification. Naturally when in a following paper we will consider the case of the trion in semiconductors, this simplification will be unacceptable since the hole mass is usually even lighter than the conduction band electronic mass. But it is easy to see that this does not bring in practice any sizeable complication. The ground state energy is known \cite{jcp} in this case 
from variational type of calculations with an extremely high precision. Keeping the precision suitable for our purpose it is given by $E_0=2.903724$ a.u.
$=5.807448$ Rydberg.

Since $m_h=\infty$
implies $r=0$ and $\mu =m_e$, Eq.(\ref{eqThom}) and Eq.(\ref{eqShom}) simplify into
\begin{eqnarray}\label{eqThominf}
T({\bf p},{\bf p}'){=}{-}\frac{2m_e}{2m_e E{+}{\bf p}^2{+}{\bf p}'^2}
\sum_{\bf k}\bigg[T_2\left({-}E{-}\frac{{\bf p}'^2}{2m_e},{\bf p},{\bf k}\right)T({\bf p}',{\bf k})
{+}T_2^e\left({-}E{-}\frac{({\bf p}+{\bf p}')^2}{4m_e},\frac{{\bf p}{-}{\bf p}'}{2},{\bf k}\right)S({\bf p}+{\bf p}',{\bf k})\bigg]
\end{eqnarray}
\begin{eqnarray}\label{eqShominf}
S({\bf Q},{\bf q}){=}{-}\frac{4m_e}{4m_e E{+}{\bf Q}^2{+}4{\bf q}^2}
\sum_{\bf k}\;T_2\left(-E{-}\frac{(\frac{{\bf Q}}{2}+{\bf q})^2}{2m_e},
(\frac{{\bf Q}}{2}{-}{\bf q}),{\bf k}\right)\;T(\frac{{\bf Q}}{2}{+}{\bf q},{\bf k})
+({\bf q}\leftrightarrow\!-{\bf q})\hspace{27mm}
\end{eqnarray}
Here we have already used explicitly the fact that $T_2(\{\Omega,{\bf P}\};{\bf k},{\bf k}')=T_2(\Omega-{\bf P}^2/2M,{\bf k},{\bf k}')$ as indicated at the beginning of
section \ref{gip}. This yields $T_2(\{-E-{\bf p}'^2/2m_e,-{\bf p}'\};{\bf p},{\bf k})=T_2(-E-{\bf p}'^2/2m_e,{\bf p},{\bf k})$ since $M=m_e+m_h$ is infinite in this case. Similarly $T_2^e(\!\{-E,{\bf p}+{\bf p}'\};({\bf p}{-}{\bf p}')/2,{\bf k})=T_2(-E-({\bf p}+{\bf p}')^2/4m_e,({\bf p}{-}{\bf p}')/2,{\bf k})$ since here $M=m_e+m_e=2m_e$.

We now make use of the expression found in section \ref{coulT} for the Coulomb $T_2(\omega,{\bf k},{\bf k}')$. It is naturally quite convenient 
to use the same reduced units as in section \ref{coulT}, namely take $a_0=4\pi \epsilon/(2\mu Z e^2)$ as unit of length, $1/(2\mu a_0^2)$ as energy unit,
setting $\omega =-\kappa^2/(2\mu a_0^2)$, and express the wavevectors in terms of the unit $1/a_0$. These reduced units are for the electron-nucleus
Coulomb problem, which means in our case that the reduced mass is merely the electronic mass $\mu=m_e $, and $Z=2$. However we have also to use the solution
of the electron-electron Coulomb problem to obtain $T_2^e$. For this case we have naturally to translate back the result of section \ref{coulT} in physical
units, and then use the above reduced units to write the proper reduced expression. Finally, as in section \ref{coulT}, we set $T_2=(a_0/2m_e) t_2$ and
similarly $T_2^e=(a_0/2m_e) t_2^e$. In the following we use the same notations as above for the wawevectors, but they have now to be understood as being
in reduced units. However in order to avoid any confusion we use small letters $t({\bf p},{\bf p}')$ and $s({\bf Q},{\bf q})$, instead of $T({\bf p},{\bf p}')$ and 
$S({\bf Q},{\bf q})$ to indicate that we work now with reduced units.

In this way, with $K^2=2m_ea_0^2 E$, Eq.(\ref{eqThominf}) and Eq.(\ref{eqShominf}) become
\begin{eqnarray}\label{eqThominfred}
t({\bf p},{\bf p}'){=}{-}\frac{1}{K^2{+}{\bf p}^2{+}{\bf p}'^2}
\int \!\!\frac{d{\bf k}}{(2\pi )^3}\bigg[t_2(\sqrt{K^2+{\bf p}'^2},{\bf p},{\bf k})\;t({\bf p}',{\bf k})
+t_2^e(2Z\kappa_e,\frac{\bf p_-}{2},{\bf k})\;s({\bf p_+},{\bf k})\bigg]
\end{eqnarray}
\begin{eqnarray}\label{eqShominfred}
s({\bf Q},{\bf q}){=}{-}\frac{2}{2 K^2{+}{\bf Q}^2{+}4{\bf q}^2}
\int \!\!\frac{d{\bf k}}{(2\pi )^3}\;t_2(\sqrt{K^2+{\bf Q_+}^2},
{\bf Q}_-,{\bf k})\;t({\bf Q}_+,{\bf k})
+({\bf q}\leftrightarrow\!-{\bf q})
\end{eqnarray}
where we have used the abbreviations ${\bf p}_{\pm}={\bf p}\pm{\bf p}'$, ${\bf Q}_{\pm}={\bf Q}/2\pm{\bf q}$ and $\kappa_e=\sqrt{K^2/2+{\bf p}_+^2/4}$,
and we have from section \ref{coulT}:
\begin{eqnarray}\label{eqt2t2e}
t_2(\kappa,{\bf q},{\bf q}')=-\frac{4\pi z}{({\bf q}-{\bf q}')^2}\int_{0}^{1}du\,\frac{u^{-\frac{1}{2\kappa}}(1-u^2)}{[u+z(1-u)^2]^2}\hspace{15mm}
t_2^e(\kappa,{\bf q},{\bf q}')=\frac{1}{Z}\frac{4\pi z_e}{({\bf q}-{\bf q}')^2}\int_{0}^{1}du\,\frac{u^{\frac{1}{2\kappa}}(1-u^2)}{[u+z_e(1-u)^2]^2}
\end{eqnarray}
with, in $t_2(\kappa,{\bf q},{\bf q}')$, $z=(\kappa^2+{\bf q}^2)(\kappa^2+{\bf q'}^2)/[4\kappa^2 ({\bf q}-{\bf q}')^2]$, and, in $t_2^e(\kappa,{\bf q},{\bf q}')$, $z_e=(\kappa_e^2+{\bf q}^2)(\kappa_e^2+{\bf q'}^2)/[4\kappa_e^2 ({\bf q}-{\bf q}')^2]$ with $\kappa_e =\kappa/(2Z )$. One can show easily from Eq.(\ref{eqThominfred}) and Eq.(\ref{eqShominfred}) themselves that 
$t({\bf p},{\bf p}')$ and $s({\bf Q},{\bf q})$ go very rapidly to zero when the modulus of any of the argument wavevectors go to infinity.
We notice also that the calculation to be performed to obtain $s({\bf Q},{\bf q})$ in Eq.(\ref{eqShominfred}) is just the same as the one appearing for the first term
in the r.h.s. of Eq.(\ref{eqThominfred}) provided the substitution ${\bf p} \to {\bf Q}_-$ and ${\bf p'} \to {\bf Q}_+$ is made. The second term of
Eq.(\ref{eqThominfred}) is also quite analogous to the first one.

From rotational invariance $t({\bf p},{\bf p}')$ depends only on the moduli $p$ and $p'$ of the two wavevectors, together with the angle $\alpha $ between them,
so we may write $t({\bf p},{\bf p}') \equiv t(p,p',\cos \alpha)$. Turning now to the practical evaluation of the first integral in the r.h.s. of Eq.(\ref{eqThominfred}), 
we see that for fixed polar angle $\theta$ of ${\bf k}$ with respect to ${\bf p'}$, only the factor $t_2(\sqrt{K^2+{\bf p}'^2},{\bf p},{\bf k})$ depends on the azimuthal
angle $\varphi$ of ${\bf k}$ with respect to ${\bf p'}$. It turns out that this integration can be performed analytical from Eq.(\ref{eqt2t2e}),
so we are left with performing the $k$ and the $\theta$ integration numerically, with in addition the $u$ integration to be performed to obtain $t_2$ and $t_2^e$.
The same point can be made for the second term in the r.h.s. of Eq.(\ref{eqThominfred}) provided we replace ${\bf p}$ and ${\bf p'}$ by ${\bf p_-}$ and ${\bf p_+}$.
Finally we have noticed that the evaluation in Eq.(\ref{eqShominfred}) is related to the one in Eq.(\ref{eqThominfred}). Hence in practice the
integral equations Eq.(\ref{eqThominfred}) and Eq.(\ref{eqShominfred}) are actually two-dimensional integral equations, which is fairly simple to handle numerically.
The integration with respect to $\theta$ is quite conveniently performed with Gaussian integration, while an appropriate Simpson method is well suited for
the $k$ integration.

There are two small problems arising in the $u$ integration, one for $u \to 0$, and the other for $u \to 1$. So it is better to split the integral into two integrals,
in order to handle separately the $u \to 0$ and $u \to 1$ problems. The $u \to 0$ problem, which results from the somewhat singular behaviour 
$u^{\mp 1/(2\kappa)}$ is easily settled by an appropriate change of variables. In the first term in the r.h.s. of Eq.(\ref{eqThominfred}), the $u \to 1$ problem arises because, when ${\bf k} \to {\bf p}$, $z$ in Eq.(\ref{eqt2t2e}) diverges and the $u$ integrand behaves as $1/(1-u)^2$, 
resulting in a divergent $u$ integral for $u \to 1$. The dominant behaviour can be extracted and handled analytically. Ultimately this
leads to an integrable logarithmic singularity at $k=p$ in the $k$ integration, which is nevertheless annoying numerically. We prefer to avoid this difficulty altogether 
by having an integrand which is exactly zero for ${\bf k}={\bf p}$. This is done by subtracting and adding a same quantity in the integrand. Explicitly
the first term in the r.h.s. of Eq.(\ref{eqThominfred}) becomes:
\begin{eqnarray}\label{}
\frac{1}{8\pi ^2 \kappa ^2}  \int \!\!d\Omega_k \int_{0}^{\infty} \!\!\!dk \left[k^2(k^2+\kappa ^2)t(p',k,\cos \theta)-p^2(p^2+\kappa ^2)t(p',p,\cos \alpha )\right]
\int_{0}^{1}\!\!du\,\frac{u^{-\frac{1}{2\kappa}}(1-u^2)}{[u({\bf p}-{\bf k})^2+z_0(1-u)^2]^2} \\ \nn
+\frac{1}{8\pi ^2 \kappa ^2}p^2(p^2+\kappa ^2)t(p',p,\cos \alpha ) \int \!\!d\Omega_k \int_{0}^{\infty} \!\!\!dk 
\int_{0}^{1}\!\!du\,\frac{u^{-\frac{1}{2\kappa}}(1-u^2)}{[u({\bf p}-{\bf k})^2+z_0(1-u)^2]^2}\hspace{30mm}
\end{eqnarray}
where, for simplicity and clarity, we have not written explicitly the analytical result of the $\varphi$ integration, in the $d\Omega_k=\sin \theta \,d\theta \,d\varphi$
integration. We have set $z_0=(\kappa^2+{\bf p}^2)(\kappa^2+{\bf k}^2)/(4\kappa^2)$. In the second term both the $k$ and the $\Omega_k$ integration
can be performed analytically, and one is only left with the $u$ integration to be performed numerically. The second term in Eq.(\ref{eqThominfred}) and the r.h.s.
in Eq.(\ref{eqShominfred}) are handled in essentially the same way.

Let us call $A$ the linear operator corresponding to the action of the r.h.s of Eq.(\ref{eqThominfred}) and Eq.(\ref{eqShominfred}) on the two-dimensional column
vector $(t,s)$. Solving Eq.(\ref{eqThominfred}) and Eq.(\ref{eqShominfred}) is equivalent to find an eigenvector of $A$ with the eigenvalue $\lambda=1$. This
problem has a solution only if $E$ corresponds to the energy of bound states of our three-body hamiltonian. In particular for very large $E$ (that is very large
binding energy), which implies large $\kappa$ and $\kappa_e$, $t_2$ reduces to the Born approximation and, from the prefactors in Eq.(\ref{eqThominfred}) 
and Eq.(\ref{eqShominfred}), the operator $A$ goes to zero and all its eigenvalues are quite small. Hence none of them can be equal to $1$, and there is
no state with very large binding energy, as expected. If we decrease $E$ the largest positive eigenvalue $\lambda_{\mathrm max}$ of $A$ will grow. 
When it reaches $1$ we will have obtained the largest possible value for $E$ corresponding to an eigenstate. 
In other words we will have the ground state energy. 

It is easy to obtain the largest
eigenvalue of $A$ by applying iteratively $A$ to some convenient starting vector $(t_0,s_0)$. Indeed iterating $n$ times is equivalent to applying the operator
$A^n$ to $(t_0,s_0)$. But for large values of $n$, $A^n$ is dominated by its largest eigenvalue $\lambda_{\mathrm max}^n$ and is essentially equivalent
to a projection on the corresponding eigenvector and multiplication by $\lambda_{\mathrm max}^n$. This allows to identify conveniently $\lambda_{\mathrm max}$
and the corresponding eigenvector. Actually this procedure works only if the spectrum of $A$ does not have nasty features, such as closely spaced largest and
second largest eingenvalues,
or large negative eigenvalues. Fortunately we have found that, in our case, this procedure works quite nicely. We have found that in practice 20 iterations
gave already a satisfactory convergence for the precision we have considered. For example going up to 40 iterations did not bring any sizeable change.
It is also convenient, in order to find the ground
state energy, to start from the situation where the electron-electron interaction is zero (for which the answer is known) and crank it progressively to its actual
value. In this way, at each stage, the range where the ground state energy lies is fairly well known.

In practice we find that Gaussian integration is extremely efficient. Typically the precision for the ground state energy increases exponentially with the number of Legendre polynomials used. For practical purposes going up to $\ell=5$ is already quite enough, although we have used $\ell=10$ in the results given below. 
The limitation for precision comes mainly from the mesh we use for the $k$ integration. With our notation we have for the ground state energy exactly $2K^2=1$ when the electrons are not interacting, while when they interact the known result \cite{jcp} is $2K^2=0.72593$. Taking successively 10 points, 20 points and 40 points for
our $k$ mesh, we have found numerically for the non-interacting electrons $2K^2=1.00742$, $2K^2=1.00066$ and $2K^2=1.00007$, 
while for the interacting electrons we
obtain $2K^2=0.73488$, $2K^2=0.72641$ and $2K^2=0.72604$. Hence we see that we obtain in both cases the exact result with a precision which is typically
$10^{-2}$, $10^{-3}$ and $10^{-4}$ successively. This is a quite clear numerical check that our method converges rapidly 
toward the exact result when the precision of
the numerical procedure is increased. On the other hand the calculation time increases markedly with improved precision. On our MacBookPro computer, the
$1\%$ precision (which is in practice quite enough for the determination of the trion binding energy) takes typically 1 mn. We need 30 mn to obtain the $10^{-3}$
precision and 280 mn for $10^{-4}$. These times are for calculations starting without a priori information on the result and no refinement. 
Naturally they are markedly shortened as soon as some information from preceding calculations are used and/or the mesh is made more precise 
only near the end of the calculation when one looks for improved precision on the result. 

\begin{figure}
\centering
{\includegraphics[width=\linewidth]{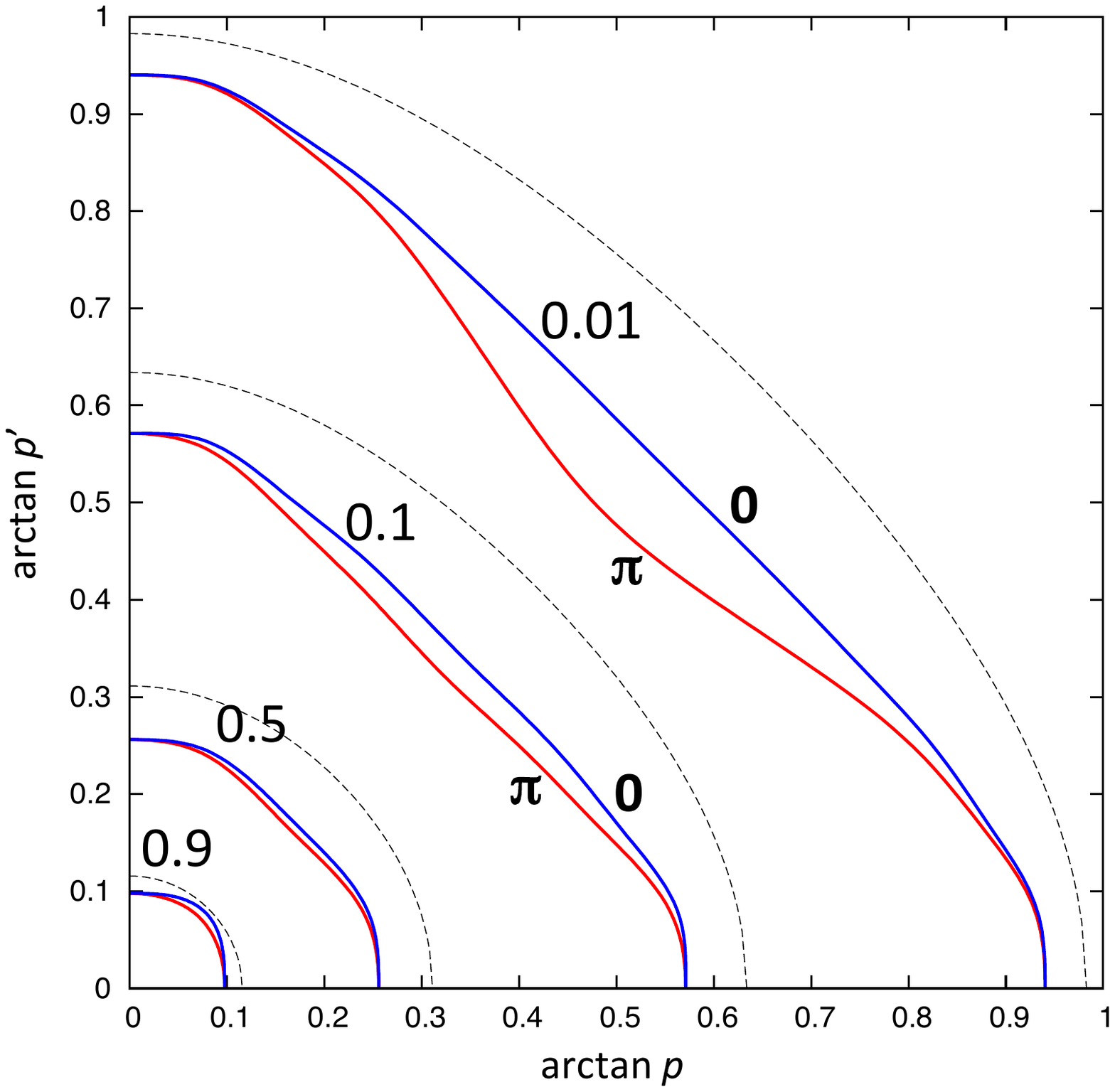}}
\caption{(Color on line) Contour lines for the two-electron wavefunction. The wavefunction is normalized to $1$ for $p=p'=0$. The contour lines correspond to
the respective values 0.9 , 0.5 , 0.1 and 0.01 for the wavefunction as indicated. The wavefunction depends also on the angle between ${\bf p}$ and ${\bf p'}$, but
the dependence is rather weak, and only the contour lines for the angles $0$ (blue) and $\pi $ (red) are drawn, as indicated for 0.1 and 0.01 . 
When the angle increases from $0$ to $\pi $ the contour lines interpolate smoothly between these two limits. 
Note that the axes are for $\arctan p$ and $\arctan p'$, not $p$ and $p'$. The dashed lines correspond, for the same values 0.9 , 0.5 , 0.1 and 0.01 
for the wavefunction, to the case where the two electrons are non-interacting, the result being obtained both analytically and numerically.}
\label{figwav}
\end{figure}

Let us now turn to the wavefunction. We have plotted in Fig.~\ref{figwav}, in the $(p,p')$ plane, the contour lines of the normalized wavefunction for values 0.9 , 0.5 , 0.1 and 0.01 , the wavefunction being normalized to 1 for $p=p'=0$. Naturally the wavefunction depends also on the angle between ${\bf p}$ and ${\bf p'}$,
but this dependence turns out to be rather weak. Hence we have plotted only the results when ${\bf p}$ and ${\bf p'}$ are parallel and antiparallel. The correlation between the two electrons is seen to be stronger in this latter case. When the angle goes from $0$ to $\pi$, the contour lines interpolate smoothly between these two limits. Naturally there is no angular dependence when ${\bf p=0}$ or ${\bf p'=0}$, so the angular dependence is strongest along the diagonal $p=p'$.
For the sake of comparison we have also plotted the same contour lines in the case where the two electrons are non interacting, and uncorrelated. In this case the wavefunction is naturally known analytically, being the product of the two single electron wavefunctions. As expected the electron-electron repulsion leads to
an expansion of the wavefunction in direct space, and correspondingly to a contraction in $k$ space. This is indeed what we find. Finally let us indicate that, in
the case of non-interacting electrons, our numerical solution for the wavefunction is in excellent agreement with the known analytical expression.

\section{Conclusion}

In this paper we have presented an exact general approach for the solution of the three-body problem for a general interaction,
which happens to be simple and fast, and applied it to the case of the Coulomb interaction. Rather than starting with the Schr\"{o}dinger equation
for this problem, it makes rather use of a corresponding integral equation derived from the consideration of the scattering properties of the system,
namely when one body is scattered by the two-body system formed by the two other ones. In this way one makes full use of the solution of the
two-body problem which appears through the corresponding T-matrix, and the interaction does not appear explicitly but only through this known T-matrix.
We have shown that the frequencies can be eliminated and only on-the-shell evaluations of the involved vertices appear. When two body have the same
interactions, finding the ground state (or any bound state) of the three-body system amounts to find for which energy two coupled 3-dimensional 
linear integral equations have a solution. The wave function is directly obtained from the corresponding solution. 

We have applied this approach to the well-known Helium atom ground
state problem, making use of the T-matrix for the Coulomb potential obtained by Schwinger. In this case the linear integral equations turn out to be
2-dimensional. We obtain a perfect numerical agreement with the known result for the ground state energy. We expect to apply this approach
in the near future to other three-body problems of interest, and in particular to the trion problem in semiconductors.

\end{document}